%% file: ch+_uves.tex
\documentclass[twocolumn,letterpaper]{aa}
\usepackage{graphicx}
\usepackage{txfonts}
\begin{document}



\newcommand{\rism}{$78.27 \pm 1.83$}
\newcommand{\scatterism}{$12.7$}
\newcommand{\nsigscat}{$6.9$}
\newcommand{\fracdev}{16.2\%~}

%


\title{ Interstellar $^{12}$C/$^{13}$C ratios through
CH$^+\lambda\lambda 3957,4232$ absorption in local clouds: incomplete
mixing in the ISM \thanks{Based on observations obtained with UVES at
the ESO Very Large Telescope, Paranal, Chile (proposal
No. 71.C-0367(A))}}


%

   \titlerunning{Interstellar $^{12}$C/$^{13}$C}


   \author{S. Casassus 
          \inst{1}
           \and O. Stahl   \inst{2}  \and T.L. Wilson\inst{3, }\inst{4}}

   \offprints{T.L. Wilson}

   \institute{ 
              Departamento de Astronom\'{\i}a, Universidad de Chile,
              Casilla 36-D, Santiago, Chile \\
              \email{simon@das.uchile.cl}
         \and
             Landessternwarte K\"onigstuhl, 69117 Heidelberg, Germany\\
              \email{O.Stahl@lsw.uni-heidelberg.de}
         \and
	     ESO, Karl-Schwarzschild-Str. 2, D-85748 Garching bei
              M\"unchen \\
              \email{twilson@eso.org}
         \and
             Max-Planck-Institut f\"ur Radioastronomie, Postfach 2024, D-53010 Bonn, Germany}


   \abstract{The $^{12}$C/$^{13}$C isotope ratio is a tracer of
stellar yields and the efficiency of mixing in the ISM.
$^{12}$CH$^{+}$/$^{13}$CH$^{+}$ is not affected by interstellar
chemistry, and is the most secure way of measuring $^{12}$C/$^{13}$C
in the diffuse ISM.  $R=^{12}$C/$^{13}$C is 90 in the solar system.
Previous measurements of $^{12}$CH$^{+}\lambda\lambda$3957.7,4232.3
and $^{13}$CH$^{+}\lambda\lambda$3958.2,4232.0 absorption toward
nearby stars indicate some variations in $^{12}$C/$^{13}$C, with
values ranging from 40 to 90 suggesting inefficient mixing. Except for
the cloud toward $\zeta$Oph, these $R$ values are strongly affected by
noise. With UVES on the VLT we have improved on the previous
interstellar $^{12}$C/$^{13}$C measurements.  The weighted
$^{12}$C/$^{13}$C ratio in the local ISM is \rism, while the weighted
dispersion of our measurements is \scatterism, giving a
\nsigscat~$\sigma$ scatter.  Thus we report on a \nsigscat~$\sigma$
detection of \fracdev root-mean-square variations in the carbon
isotopic ratio on scales of $\sim$100~pc: $R= 74.7 \pm 2.3$ in the
$\zeta$Oph cloud, while $R = 88.6 \pm 3.0$ toward HD152235 in the
Lupus clouds, $R = 62.2 \pm 5.3$ towards HD110432 in the Coalsack, and
$R = 98.9 \pm 10.1$ toward HD170740. The observed variations in
$^{13}$C/$^{12}$C are the first significant detection of chemical
heterogeneity in the local ISM.

 }

   \maketitle
%

\section{Introduction}


The ratio of the $^{12}$C to $^{13}$C isotope is a good tracer of the
amount of stellar processing in low and intermediate mass stars during
the asymptotic giant branch (AGB) phase.  The third dredge-up
increases $^{12}$C/$^{13}$C to about $\sim300$, while stars massive
enough to undergo ``hot bottom burning'' bring their surface carbon
isotopic ratio to CNO processing equilibrium,
$^{12}$C/$^{13}$C$\sim$3. The carbon yields of AGB stars are sensitive
functions of their initial masses and metallicities (e.g. Renzini \&
Voli \cite{ren81}, or Casassus \& Roche \cite{cas01} for a population
synthesis approach and references to model results).

A measurement of the $^{12}$C/$^{13}$C ratio in the interstellar
medium (ISM) gives important data for the total amount of low mass
stellar evolution and subsequent enrichment of the ISM in the
Galaxy. The terrestrial value of $^{12}$C/$^{13}$C is 90 (Rosman \&
Taylor \cite{ros98}).  The $^{12}$C/$^{13}$C ratio is a cornerstone of
models of the nuclear history of our galactic ISM, that is, the
Galactic Chemical Evolution.  Models of chemical evolution predict a
decrease in the $^{12}$C/$^{13}$C ratio with time for a given
galactocentric distance, and a decrease with galactocentric distance
in the Galaxy (e.g. Palla et al. \cite{pal00}, their Fig.~4 and 5.).
Such models use the assumption that mixing in the ISM is complete and
restricted to material at a given galactocentric distance (azimuthal
mixing). A comparison of results between our Galaxy and other galaxies
will give important data for the nuclear processing history (see,
e.g., Tosi 2000, Prantzos 2001).

There are a large number of measurements of the $^{12}$C/$^{13}$C
ratio from radio astronomy data (see, e.g., Wilson \& Rood
\cite{wil94}). However, these results may be affected by interstellar
chemistry in two ways. These are chemical fractionation (which
enriches molecules in $^{13}$C and thus lowers the ratio) and
selective dissociation (which destroys the rarer species more, and
thus raises the ratio). The $^{13}$C enrichment is due to
fractionation of the CO molecule via the ion-molecule process,
\begin{equation}
  ^{13}\mathrm{C}^{+} + ~^{12}\mathrm{CO}
\stackrel{k}{\rightleftharpoons} ~^{12}\mathrm{C}^{+}
+ ~^{13}\mathrm{CO} + \Delta E,
\end{equation}
with $\Delta E / k = 35~$K and $k = 2~10^{-10}~$cm$^3$~s$^{-1}$ in
both directions (Watson et al.  \cite{wat76}).  If the reaction
reaches equilibrium, 
\begin{equation} \label{eq:co_eq}
  \frac{n(^{13}\mathrm{C}^{+}) ~n(^{12}\mathrm{CO})}{ n(^{12}\mathrm{C}^{+})
  ~n(^{13}\mathrm{CO})} = \exp\left(
  -\frac{\Delta E}{kT} \right). 
\end{equation}
In cold gas, with $T \la 35~$K, CO will be enriched in $^{13}$C
compared to C$^{+}$.  The effect is observed in CO absorption towards
$\zeta$Oph (Sheffer et al. \cite{she92}), $\rho$Oph and $\chi$Oph
(Federman et al. \cite{fed03}).  Thus molecules whose synthesis
involve CO at a later stage than C$^{+}$ will show the CO
fractionation, and {\em vice-versa} (e.g. Watson \cite{wat78}). The
corresponding $^{13}$C depletion in C$^{+}$ is diluted in environments
where $n($C$^{+} ) \gg n($CO), such as diffuse clouds or
photo-dissocitation regions (PDRs) at $A_\mathrm{V} < 2$ (e.g.  Keene
et al. \cite{kee98}, their Fig.~2).

Ultra-high-resolution spectroscopy of CH$^{+}$ (Crawford et
al. \cite{cra94}) absorption towards $\zeta$Oph reveals broader
profiles compared to CN and CH, with an upper limit kinetic
temperature of $\sim$2000~K, and with no velocity offset. This is
consistent with CH$^{+}$ production through the endothermic reaction
\begin{equation} \label{eq:ch_prod}
\mathrm{C}^{+} + \mathrm{H}_2 + 0.4~\mathrm{eV} \rightarrow \mathrm{CH}^{+} +
\mathrm{H}.
\end{equation}
The analog of Eq.~\ref{eq:co_eq} shows that temperatures $> 1000~$K
are required for CH$^{+}$ production via Eq.~\ref{eq:ch_prod}.
CH$^{+}$ thus arises in the warm halo, or photo-dissociation region,
surrounding the $\zeta$Oph cloud.  Crawford (\cite{cra95}) and Crane
et al. (\cite{cran95}) extend the $\zeta$Oph results to five and
twenty other lines of sight through translucent clouds (i.e. with
$A_\mathrm{V} < 2$). With such a kinetic temperature, it is clear
that CO fractionation in the sites of CH$^+$ production is negligible.

The carbon isotope ratio as measured by
$^{12}$CH$^{+}$/$^{13}$CH$^{+}$ is not affected by selective
dissociation. Although photodissociation of CH$^+$ contributes to PDR
chemical networks (Sternberg \& Dalgarno \cite{ste92}), CH$^+$ is
optically thin to photodissociating UV radiation in translucent
clouds.  The strongest optical CH$^+$ line in $\zeta$Oph is $A ^1\Pi -
X ^1\Sigma^+$ at 4232~{\AA} (Morton \cite{mor75}), which reaches
maximum opacities of $\sim$0.3 (e.g. this work). Since no ultra-violet
CH$^+$ lines have been reported, they are bound to be faint, and thus
thin. The photodissociation cross-section $\sigma(\nu)$ calculated by
Kirby et al. (\cite{kir80}) is mostly due to transitions from the
ground $X^1\Sigma^+$ state to the vibrational continua of the
$2^1\Sigma^+$, $3^1\Sigma^+$, and $2^1\Pi$ states, which smooths out
any difference between isotopes. Some resonant absorption derives from
the calculation of $\sigma(\lambda)$ at wavelengths of 1418\AA,
1453\AA, and 1490\AA. But Kirby et al. (\cite{kir80}) assign oscillator
strengths of less than $10^{-3}$ in any of these peaks, which
represents Einstein $B$ absorption rates one order of magnitude
smaller than at 4232~{\AA}.



The $^{12}$CH$^{+}$/$^{13}$CH$^{+}$ ratio is thus a secure
representation of $^{12}$C/$^{13}$C in translucent clouds.  In
addition, the absorption line from $^{12}$CH$^{+}$ at 4232{{\AA}} is
separated from the line of $^{13}$CH$^{+}$ by 0.265{{\AA}}, which
permits unblended measurements.  These ratios, obtained from optical
measurements, are restricted to regions within $\sim$2~kpc of the Sun
because of the need for bright background stars, and of narrow
velocity profiles in the intervening cloud.  Eight ratios have been
measured but only one line of sight, toward {$\zeta$}Oph, has an
excellent signal to noise ratio. The average $^{12}$C/$^{13}$C ratio
from CH$^{+}$ data agrees well with those obtained for sources near
the Sun, from radio astronomy data.  This gives one confidence in the
ratios obtained from radio data.

However, the CH$^{+}$ measurements so far resulted in isotopic ratios
with a large scatter between different lines of sight, and there is
only one line of sight with an excellent signal to noise ratio.  The
previous data from CH$^{+}$ were taken with the ESO CAT, and 4-m or
smaller aperture telescopes.  The best results from the point of view
of the signal to noise ratio and the reliability of the ratio for
{$\zeta$}Oph were presented by 3 different groups (Stahl et
al. \cite{sta89}, Stahl \& Wilson \cite{sta92}, Crane et
al. \cite{cra91}, Hawkins et al. \cite{haw85}, \cite{haw93}).  The
final ratio obtained by each group was 67$\pm$5.  The {$\zeta$}Oph
value is by far the best measurement of the $^{12}$C/$^{13}$C ratio
for the ISM near the Sun.  However, the measurements for 7 other
nearby weak sources were of lower quality. The ratios for 3 of these
regions, toward HD26676 (64$\pm 6$, Centurion \& Vladilo
\cite{cen91}), HD110432 ($71\pm11$, Centurion et al. \cite{cen95}),
and $\mu$Norma ($67\pm 6$, Centurion \& Vladilo \cite{cen91} ) are in
very good agreement with the {$\zeta$}Oph ratio. But there are
discordant ratios: $R=38\pm12$ toward HD157038 (Hawkins \& Meyer
1989), and $R=49\pm15$ toward {$\zeta$}Per (Hawkins et
al. \cite{haw93}) while the ratio toward HD152235 is $R=126\pm29$ and
is $R=98\pm19$ towards HD152424 (Vladilo et al. \cite{vla93}).  Most
of the measurements are limited by noise, but the profile toward
HD152424 is complex and thus a determination of the ratio is not easy,
because of CH$^{+}$ and $^{13}$CH$^{+}$ line blending.

We repeated the CH$^{+}$ observations toward $\zeta$Oph, HD110432,
HD152235, HD152424 and HD157038, and included the new lines of sight
toward HD152236, HD154368, HD161056, HD169454, HD170740. Our goal was
first to determine whether the ratios are truly different from the
value for $\zeta$Oph, which would indicate that the mixing of the ISM
is not very fast and complete, and second, we wanted to increase the
sample of lines of sight with good to excellent signal to noise
ratios, and from this determine an average ratio near the Sun truly
representative for the local ISM.

Section~\ref{sec:obs} describes the observations, and
Section~\ref{sec:model} gives details on our procedure to measure the
isotopic ratio $R$. Section~\ref{sec:results} contains our results ,
which are then summarised and discussed in
Section~\ref{sec:summary}. Section~\ref{sec:conc} concludes.

\section{Observations and reduction} \label{sec:obs}

There are two systems of CH$^{+}$ lines which have appreciable
oscillator strengths.  Because of atmospheric absorption, the lines at
4232 \AA\ are easier to measure. The $^{13}$CH$^{+}$ line is 0.26 \AA\
($\Delta v = 19~$km~s$^{-1}$) shortward of the $^{12}$CH$^{+}$
line. The other system, at 3957 \AA, is more difficult to measure, but
has the property that the $^{13}$CH$^{+}$ line is on the long
wavelength side of the $^{12}$CH$^{+}$ line at a separation of 0.44
\AA ($\Delta v = -33~$km~s$^{-1}$).  This property has been used by
Stahl \& Wilson (\cite{sta92}) to check that there is no accidental
overlap of a weak velocity component of $^{12}$CH$^{+}$ at $-$18.8 km
sec$^{-1}$ from the deepest $^{12}$CH$^{+}$ absorption line with the
$^{13}$CH$^{+}$ line.  Nearly all previous data were taken at 4232
\AA.

Our observations were obtained with the UVES echelle spectrograph at
the VLT unit telescope Kueyen at Cerro Paranal, Chile, in three nights
between June 14/15 and June 16/17, 2003. UVES allows measuring both
line systems in the same detector setting, thus providing a means to
correct for line blending if it is apparent in the spectra.

The observations are difficult because of the need for both very high
S/N and very high spectral resolution to acquire the profile of faint
$^{13}$CH$^{+}$, against a very bright stellar continuum and close to
the much stronger $^{12}$CH$^+$ component. We need S/N ratios of at
least $\sim$10 on $^{13}$CH$^{+}$ for accurate profile fitting.  The
project requires the highest possible spectral resolution.  We
therefore used an image slicer to minimize flux losses. In addition,
the image slicer distributes the light along the slit, which improves
flat-fielding and allows for our bright targets longer integration
times before saturation occurs. UVES slicer \#2 was used, which
reformats an entrance opening of 1\farcs8 $\times$ 2\farcs2 to a slit
of 0\farcs44 width and a length of 7\farcs9, which is imaged on the
spectrograph entrance slit of 0\farcs45 width. The spectral resolution
in this configuration is $\lambda$/$\Delta\lambda$ = 75\,000. The
central wavelength was set to 4\,370 \AA, which gives a spectral
coverage from 3\,730 to 5\,000 \AA.  During our observations the
seeing was typically too good to fill the entrance aperture of the
images slicer. Therefore, for most spectra, only one or two slices
contained most of the signal, which unfortunately decreases the
expected gain of the image slicer.

Accurate flat-fielding is also important. Therefore a large number of
flat-fields (200) was obtained during daytime distributed along the
observing run. 

\begin{figure}
  \centering
  \resizebox{9cm}{!}{\includegraphics{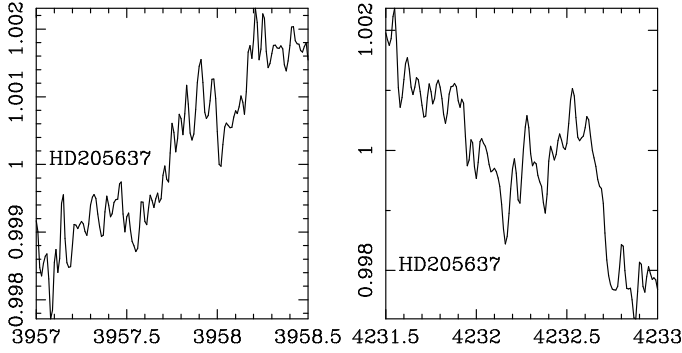}}
  \caption{Spectral region about the CH$^+$ absorption towards the
  template star HD205637, emphasizing the absence of telluric features
  at airmasses of 1.05--1.08. }
  \label{fig:template} 
\end{figure}

In addition, a rapidly rotating unreddened early-type star
($\varepsilon$~Cap = HD205637, spectral type B3V) was observed in
order to check for possible faint terrestrial features. We confirm
there are no detectable telluric features under the CH$^+$ absorption,
and show in Fig.~\ref{fig:template} the template star spectrum in the
region of interest, at airmasses $< 1.08$. Higher airmasses result in
telluric absorption features, as in the case of $\zeta$Oph discussed
below.

The brighter targets where exposed until about 50\% of the maximum
level allowed by the CCD detector was reached. For $\zeta$~Oph, this
limits the exposure time to about 5 sec, and typically a few minutes
for the fainter targets.  Series of up to 50 exposures per night were
obtained to build up the required S/N-ratio. Some targets turned out
to show too complex line profiles. These were dropped from our initial
target list.  The observations are summarized in
Table~\ref{table:obs}, which also lists the mean airmass of the target
objects in the different nights. At the higher airmass values, the
range in airmass differed from the mean by about $\pm$0.1 during the
observations, and less at smaller airmass.

We also list in Table~\ref{table:obs} the radial velocities relative
to the solar barycentre of the CH$^+$ absorption. These are calculated
from the shift in wavelength between the rest wavelengths of the
transitions and of the average of the Gaussian centroids weighted by
their equivalent widths. The uncertainty on the derived velocities is
conservatively $\sim 1$km~s$^{-1}$, and depends on the accuracy of the
rest wavelengths.

We used the Midas package originally developed for the ESO Feros
spectrograph for the reduction of the spectra (Stahl et al.,
\cite{sta99}). In order to maximize the S/N-ratio of the extracted
spectra, all flat-fields obtained during the run were averaged. After
background subtraction and flat-fielding, the spectra were extracted
with a very long slit, extracting all slices together. The wavelength
calibration of the extracted spectra was done with a 2D-polynomial,
fitting all echelle orders in one step. A mean ThAr-spectrum obtained
during day time was used for the calibration of all spectra obtained
in one night.  Finally, all spectra were merged to a 1D-spectrum and
all observations of each night averaged to a nightly mean
spectrum. Wavelengths are reported in air and refererred to the solar
barycentre.

%

%


\begin{table}[h]
\caption{Summary of the observations, giving the total number of
exposures, $N$, the integrated exposure time in seconds, and the
observed heliocentric velocities of the absorbing clouds,
$V_{\odot}$.}
\begin{tabular}{lrrrrrr}
\hline
object       & \multicolumn{3}{c}{air mass} & $N$  &  exp.    &     $V_{\odot}$     \\
             &   14/06 &  15/06 &  16/06      &         &     time   &  km~s$^{-1}$    \\
\hline
$\zeta$~Oph  & 1.16 &   1.81 &   1.71 &150    & 750     &-15.1   \\
HD110432     & 1.31 &   1.29 &   1.29 &72     & 6\,770  &+5.1     \\
HD152235     & 1.07 &   1.31 &   1.24 &65     & 16\,250 &-4.3    \\
HD152236     & -    &   1.14 &   -    &10     & 600     &-6.9    \\
HD152424     & 1.05 &   -    &   -    &10     & 2\,500  &-7.6    \\
HD154368     & 1.05 &   1.06 &   1.05 &50     & 9\,850  &-6.0    \\
HD157038     & 1.11 &   -    &   -    &10     & 2\,500  &-14.7   \\
HD161056     & 1.25 &   1.06 &   1.06 &59     & 11\,110 &-11.4    \\
HD169454     & 1.30 &   1.12 &   1.12 &50     & 22\,900 &-10.5   \\
HD170740     & 1.05 &   1.62 &   1.60 &65     & 7\,800  &-11.0   \\
HD205637     & 1.05 &   1.08 &   1.08 &150    & 3\,385  &  --\\
\end{tabular}
\label{table:obs} 
\end{table}

The nightly spectra were combined in a weighted average to produce
coadded spectra. The weights were taken as 1/$\sigma^2$, where
$\sigma$ is the noise in each spectra, as calculated from the
root-mean-square deviations from a linear fit to a region of the
spectrum devoid of conspicuous features (we chose 4204.2\AA~ to
4205.9\AA). Table~\ref{table:weights} summarise the resulting weights.


\begin{table}[h]
\caption{Relative weights used to combined the nightly spectra}
\begin{tabular}{lrrrrr}
\hline
             &  June 14 & June 15 & June 16       \\
\hline
$\zeta$Oph &     0.45 &    0.30 &     0.26 \\ 
HD110432 &     0.26 &    0.37 &     0.36 \\ 
HD152235 &     0.41 &    0.31 &     0.28 \\ 
HD154368 &     0.35 &    0.39 &     0.26 \\ 
HD161056 &     0.31 &    0.41 &     0.29 \\ 
HD169454 &     0.37 &    0.34 &     0.29 \\ 
HD170740 &     0.31 &    0.40 &     0.29 \\ 
HD205367 &     0.41 &    0.32 &     0.27 \\ 
\end{tabular}
\label{table:weights} 
\end{table}

\section{Model line profiles} \label{sec:model}

In order to use the fact that both $^{12}$CH$^{+}$ and $^{13}$CH$^{+}$
lines have the same opacity profile $\tau(v)$, we must fit the
$^{12}$CH$^{+}$ absorption with a parametrised model and scale it to
$^{13}$CH$^{+}$.

We considered using Voigt profiles to account for the intrinsic line
profiles of the CH$^{+}$ vibronic lines. In order to fit the low-level
broad wings (e.g. Stahl et al. \cite{sta89}) towards $\zeta$~Oph with
an hypothetical Lorentzian core, Einstein $A$ values of order
$10^{9}$~s$^{-1}$ are required.  Gredel. et al. (\cite{gre93}) quote
an oscillator strength of $f_\circ = 0.00545$, or $A = 2.258~10^{-4}$,
for $^{12}$CH$^{+}\lambda$4232 $\Pi-\Sigma(0,0)$ and $f_\circ =
0.00331$, or $A = 1.568~10^{-4}$, for $^{12}$CH$^{+}\lambda$3957
$\Pi-\Sigma(1,0)$. We therefore used simple Gaussian profiles to
describe the lines.

We preferred to fit the two overtones separately, and thus obtain
independent measurements with which to assess the role of
systematics. A simultaneous fit of both overtones could have helped
constrain the opacity profiles, which in our separate fits do not
always share the same Gaussian components.

%

The fitting algorithm is as follows.
\begin{enumerate}

\item Extract a 1.5{\AA} spectrum $F(\lambda)$ centred on the CH$^{+}$
line, $\lambda_1 < \lambda < \lambda_2$.

\item Define a baseline $F_c(\lambda)$ with a $4-17^\mathrm{th}$ order
Legendre polynomial to the data, as in Sembach \& Savage
(\cite{sem92}), $F_c(\lambda) = \sum^{l}_{i=1} a_i
P_i[(\lambda-\lambda_1)/(\lambda_2-\lambda_1)]$.

\item Fit for the $\{a_i\}_{i=1}^{l}$ coefficients in a least square
  sense, ignoring the neighbourhood of the ISM and telluric lines, and
  weighting the data to improve the quality of the baselines near the
  interstellar absorption lines.  Store the rms dispersion of the
  residuals as the spectrum's noise, $\sigma_F$.  $\sigma_F$ is
  updated after fitting the parametrised model; it is replaced by the
  dispersion of the residuals in
  Step~\ref{it:noiseupdate}. \label{it:local}

%

%
%

\item Define a model spectrum with
  \begin{equation}
    F^\star_m(\lambda) = F_{c}(\lambda) \exp(-\tau(\lambda)),
  \end{equation}
  where the line absorption opacity $\tau$ is a superposition of
  $n_\mathrm{g}=1$ to $6$ Gaussians on each isotope, with
  $^{13}$CH$^{+}$ components sharing the parameters of the
  $^{12}$CH$^{+}$ components, except their centroids are translated by
  a fixed velocity shift $\Delta v_\mathrm{iso}$, and their amplitude
  are scaled by a factor $f=^{13}$CH$^{+}/^{12}$CH$^{+}$, which is a
  free parameter in our fit:
  \begin{eqnarray}
    \lefteqn{    
  \tau(\lambda) = \sum_{i=1}^{n_\mathrm{g}} \left[ \tau^{\circ}_i \exp\left(
  -0.5 (\lambda - \lambda^{\circ}_i)^2 / {\sigma^{\circ}}_i^2 \right)
  + \right. }   \nonumber \\ & & \left.  ~f~
  \tau^{\circ}_i \exp\left( -0.5 (\lambda - \lambda^{\circ}_i (1+\Delta
  v_\mathrm{iso}/c))^2 /  {\sigma^{\circ}}_i^2  \right) \right].
  \end{eqnarray} \label{it:model}
  The value of $\Delta v_\mathrm{iso}$ depends on the overtone and is
  adjusted iteratively on $\zeta$Oph and checked for consistency on
  all targets.  We use $\Delta v_\mathrm{iso}(3957) =
  32.721~$km~s$^{-1}$, $\Delta v_\mathrm{iso}(4232) =
  -18.967~$km~s$^{-1}$, with uncertainties of $\sim$0.4~km~s$^{-1}$
  (roughly 1/10 the resolution element). 

\item Convolve the model spectrum with the instrument response,
  $B_\lambda$:
\begin{equation}
F_m(\lambda)  = \int d\lambda^\prime F^\star_m(\lambda^\prime) B(\lambda - \lambda_\prime).
\end{equation}
We estimate the instrumental response from the arc lamp spectrum. The
smallest line-widths are indicative of resolving powers of $\approx
83\,000\pm 2000$ rather than the nominal $75\,000$ for our setup.  The
instrumental response is approximated to a Gaussian, with a FWHM given
by $\lambda_\circ /83000$, where $\lambda_\circ$ is the center
wavelength of each CH$^+$ overtone.

\item Calculate the goodness of fit function
\begin{equation} \label{eq:chi2}
\chi^2 = \sum_j (F(\lambda_j)-F_m(\lambda_j))^2/\sigma^2_F,
\end{equation} using the  noise from Step~\ref{it:local}. We use a
single noise value and neglect possible variations of the noise level,
as would be the case for Poisson statistics. The noise variations are
important for very steep spectra, which is not our case, or under deep
absorption lines. But the accuracy of the model line profiles under
$^{12}$CH$^+$ is generally not important to infer isotope ratios (only
the wings matter).

\item Optimize the model parameters by minimizing $\chi^2$.  Perform
  an initial estimate of the global minimum with a genetic algorithm
  ({\tt pikaia}, Charbonneau \cite{cha95}), and optimize with the
  downhill simplex method ({\tt amoeba}, Press et
  al. \cite{pre86}). \label{it:opt}

\item Update the noise of the stellar spectrum by replacing $\sigma_F$
  with the rms dispersion of the residuals (the difference between the
  observed and model spectra).  \label{it:noiseupdate}

%
%

\item Estimate the uncertainty in model parameters by two methods:
\begin{enumerate}
\item search parameter space one parameter at a time to find the
  region enclosing 68.3\% confidence for the $\chi^2$ distribution
  with one degree of freedom, thus obtaining different upwards and
  downwards error bars and accounting for statistical bias,
\item estimate $1\sigma$ uncertainties from the curvature matrix of $\chi^2$
  (i.e. approximate to normal errors). 
\end{enumerate} 
\end{enumerate}

The UVES spectra are oversampled, so that the spectral datapoints are
correlated.  Independent datapoints are separated by roughly one
resolution element, of $\Delta \lambda = \lambda/75000 \approx
0.05~{\AA}$. Since the spectra are sampled with an interval of
0.01~{\AA}, $N_\mathrm{cor}\sim5$ consecutive datapoints are
correlated. In some cases the number of Gaussian components involves
as many free-parameters as independent data points (as for HD154368
and HD169454). Some Gaussian components may thus be redundant. But we
caution that our purpose is to obtain isotopic ratios, not to report
on the velocity structure of the ISM in CH$^+$. To constrain the
opacity profile consistently we would have had to fit the two
overtones simultaneously.

The correlation between the spectral datapoints is ignored when
fitting for the parametrised model. The off-diagonal terms of the
covariance matrix add terms to the $\chi^2$ goodness-of-fit estimator
(Eq.~\ref{eq:chi2}). The number of additional terms is roughly
$N_\mathrm{cor}$ per diagonal element, which increase $\chi^2$ by a
factor $N_\mathrm{cor}$. But the noise estimate from the dispersion of
the residuals also increases due to the correlation of the datapoints,
the actual noise should be $\sigma_F \times \sqrt{N_\mathrm{cor}}$. Both
corrections cancel out and leave $\chi^2$ as in Eq.~\ref{eq:chi2}.

\begin{figure}
  \centering
  \resizebox{9cm}{!}{\includegraphics{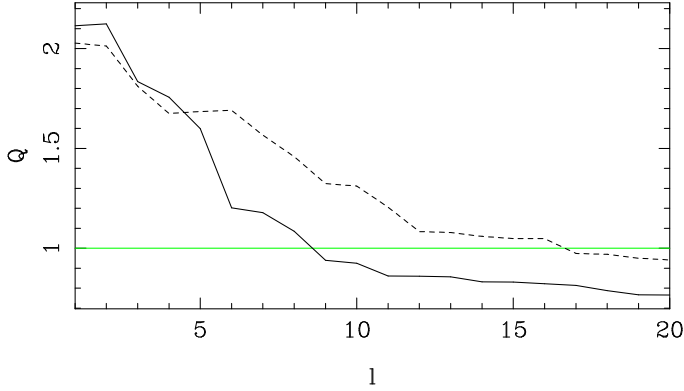}}
  \caption{The ratio of measured to expected noise, $Q$, as a function
  of the order of the Legendre polynomial used to define the
  baselines, $l$. The data are for 25 exposures of HD170740 on June
  15$^\mathrm{th}$, for two 1.5~$\AA$ spectral regions centred on
  $\lambda$4232 (solid line), and on $\lambda$3957 (dotted line).}
  \label{fig:Q} 
\end{figure}

We tested for systematic noise by comparing the noise in the nightly
coadded spectra, $\sigma_\mathrm{coadd}$, with that expected from the
noise in each of the $N_\mathrm{exp}$ individual exposures,
$\sigma_1$. For thermal noise it should hold that
$\sigma_\mathrm{coadd} = \sigma_1 / \sqrt{N_\mathrm{exp}}$. We define
$Q = \sigma_1 / (\sqrt{N_\mathrm{exp}} \sigma_\mathrm{coadd})$, and
raise the order of the baselines until $Q \sim 1$. The tests were run
for the case of HD170740, which presents the most ragged
continuum. The reference baseline is defined on the coadded spectrum,
and the noise is given by the dispersion of the residuals, excluding
the spectral regions affected by the ISM lines (these regions are
defined on Fig.~\ref{fig:HD170740}). We scale the reference baseline
to the line-free continua of single exposures using the ratio of their
median values.  Fig.~\ref{fig:Q} shows we need to reach orders of $l
\sim$15 for $\lambda$3957, and of $l \sim$8 for $\lambda$4232. Below
these orders the continua are affected by systematic
noise. Unfortunately the level of systematic noise changes for each
observation and setting, so that we can only use the tests on HD170740
as guidelines for the other lines of sight. In practice we chose the
smallest order that is compatible with the data, without exceeding the
limits in HD170740.

The isotopic ratio derived from the fits is the ratio of the column
densities of each isotope,
$R=N(^{12}\mathrm{CH}^+)/N(^{13}\mathrm{CH}^+)$. With the best fit $f$
we can calculate $R=1/f$ for a line of sight with constant $R$.  We
neglect the small difference in oscillator strength between
$^{12}$CH$^+$ and $^{13}$CH$^+$.  The two ions have slightly different
vibrational structures which will lead to small changes in the
$f_\circ$ values for individual vibrationally resolved transitions
(J.~Tennyson, private communication). We are thus neglecting
$f_\circ$-value differences of the order of the relative difference in
reduced mass, or $\Delta f_\circ / f_\circ \sim~1/170$.

In taking statistics on the values of $R=1/f$ in the local ISM care
must be taken to assess the significance of measurements along
individual sightlines.  The uncertainties on individual fits do not
include the systematic error involved in baseline definition.  If
baselines were known {\em a priori}, the expectation value and
uncertainty on $f$ for each individual sightline would be obtained
from the statistics of multiple measurements of the same
quantity. Thus, in the absence of this systematic error, $\langle f
\rangle$ and $\sigma(f)$ correspond to the weighted average and
quadratic sum of the weighted uncertainties, where the weights are
taken as the inverse variance of each measurement, $1/\sigma^2$:
\begin{eqnarray}
\langle f \rangle & =  & \sum_{i=1}^{N} f_i w_i,  \\
 \sigma^2_1(f)  & =  &  \sum_{i=1}^{N} (\sigma_i~ w_i)^2  = 1/\sum_{i=1}^{N} (1/\sigma_i)^2,
\end{eqnarray}
for $N$ measurements on a given sightline, and where the individual
errors are those derived from the curvature matrix. 

To estimate the systematic error involved in baseline definition, we
compare {\em a posteriori} the scatter of individual measurements with
that expected from the noise level. We use the rms dispersion
\begin{equation}
 \sigma_2(f)   =   \sqrt{ \sum_i  w_i~  \left[ (f_i  - \langle f \rangle) \right]^2}.
\end{equation}  
The uncertainty $\sigma_1(f)$ neglects the systematic errors.
$\sigma_2(f)$ uncertainties will give larger error bars, but are
probably too conservative considering we are in fact repeatedly
measuring the same quantity. Thus the actual uncertainty on our
measurement of $f$ for each sightline is intermediate between
$\sigma_1(f)$ and $\sigma_2(f)$. For this reason we list both
values. We have set $\sigma_2 = \sigma_1 $ if $\sigma_2 < \sigma_1 $,
as is the case when only one ratio is available to sample the effects
of systematic uncertainties in the continuum definition (or when one
ratio has much higher weight than the others).

\section{Results} \label{sec:results}

An application of the fitting procedure described in
Sec.~\ref{sec:model} to our sample of 10 lines of sight gives the best
fit parameters and isotopic ratios listed in Table~\ref{table:all}.

The isotopic ratio $f=^{13}$C$/^{12}$C is better suited as free
parameter than $R=1/f$ for the purpose of fitting individual spectra,
and subsequently averaging the best fit values. For noisy data the
error propagation when using $R$ would involve second order expansions
in $R(f)$ when calculating averages and dispersions. In what follows
we prefer to list our results in terms of $f$, and revert to the more
common usage of $R$ in Section~\ref{sec:summary} when comparing with
previous data.

Table~\ref{table:all} lists the isotope ratios and best fit
parameters.  It is apparent the $\lambda$4232 line generally gives the
best estimate of $R$. We note from Table~\ref{table:all} that the
normal errors are good approximations to the 68.3\% confidence limits,
which justifies their subsequent use in the weighted averages. We also
note the weighted average of several measurements for the same line of
sight are consistent, within the errors, with the fit to the coadded
spectra (which are in general excluded from the averages), giving
confidence in our procedure.

 
\begin{table*}
\caption{We list $f\times 10000$ to improve the clarity of the
  table. $\sigma(f)$ refers to the Gaussian errors (those derived from
  the $\chi^2$ curvature matrix).  The average values are quoted as
  $\langle f \rangle \pm \sigma_1 (\sigma_2)$ (see text for details),
  and exclude the coadded spectra. Equivalent widths for
  $^{12}$CH$^+$, $W_\lambda$, are given in m\AA. The stars select the
  measurements included in the combined values of $R$; the selection
  are discussed in the text.  
    \label{table:all}  } 
\begin{center}
\input{table_ALL_paper.tex}
\end{center}
\end{table*}

\subsection{Notes on individual objects}

\subsubsection{$\zeta$~Oph}

\begin{figure*}
  \centering
  \includegraphics{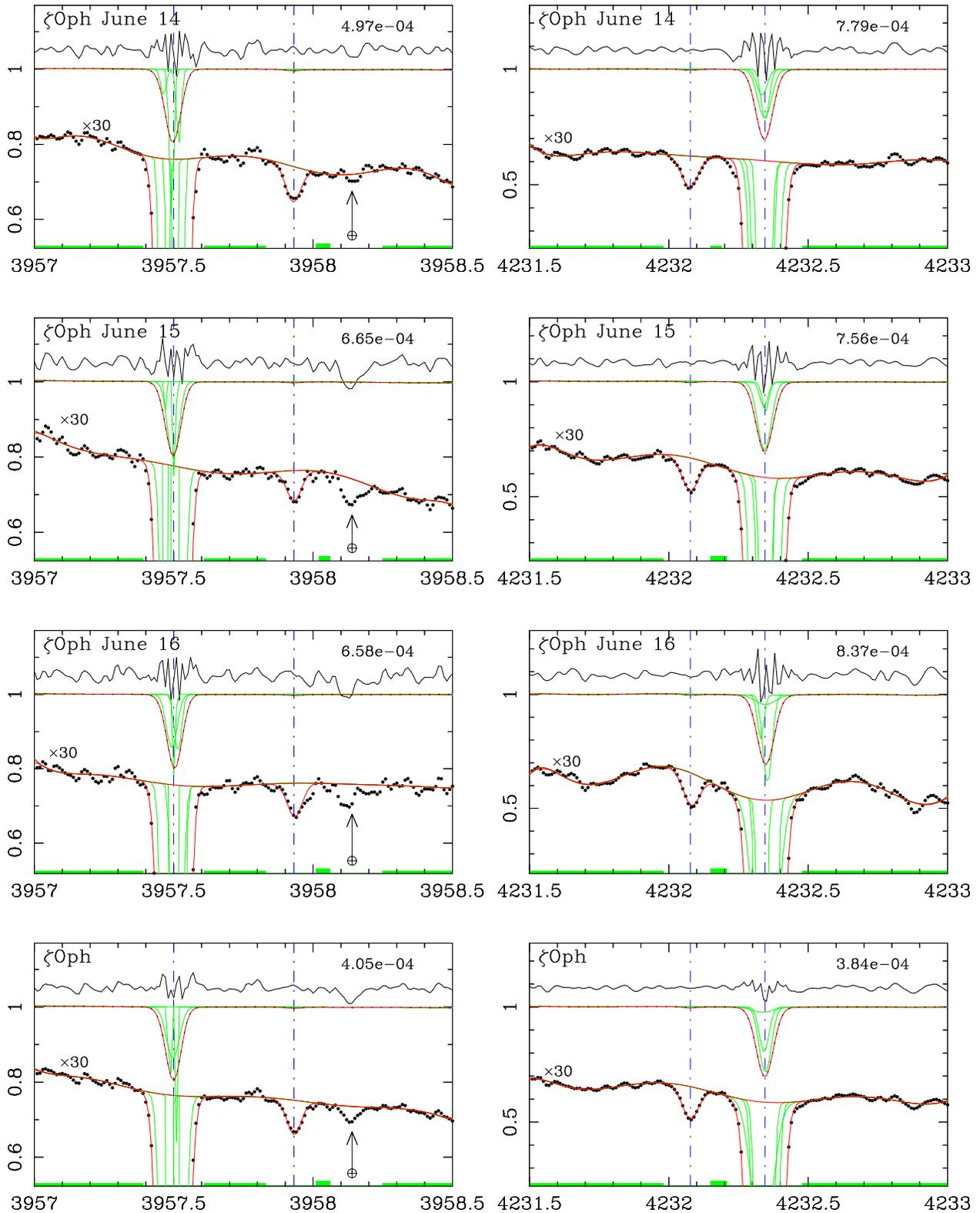}
  \caption{$\zeta$Oph spectra. Wavelengths are in Angstroms, and flux
  densities in arbitrary units. The individual Gaussian components
  comprising the fit, prior to folding with the instrumental response,
  are shown only on $^{12}$CH$^{+}$ as light gray lines, but are
  omitted from $^{13}$CH$^{+}$ for clarity (see text for details). The
  vertical dashed lines mark the line centroids, at the average of the
  Gaussian centroids weighted by their equivalent widths. The units of
  the $y-$axis are arbitrary, and are scaled so that the median of the
  object spectrum is unity.  Also shown is a magnified version of the
  object spectrum, by the factor indicated on the figure, and offset
  for clarity.  The residual spectrum is shown on top of the object
  spectrum, and is also magnified by the factor indicated on figure.
  The noise used to compute the significance of the fits is labelled
  on the top right. The noise is the rms dispersion of the residual
  spectrum, and is in the same units as the $y$-axis. The height of
  the shaded rectangles on top of the $x$-axis indicates the relative
  weights used in the baseline definition.}
  \label{fig:zoph} 
\end{figure*}

From {\em Hipparcos} results, the distance to $\zeta$Oph gives an
upper limit to the absorbing cloud of 140$\pm$14~pc.  The spectra of
$\zeta$Oph around the CH$^{+}$ lines are shown on Fig.~\ref{fig:zoph},
where we have also plotted the Gaussian components on the
$^{12}$CH$^{+}$ line (the individual components on the $^{13}$CH$^{+}$
line are left out for clarity).

Residual features under the absorption lines vary on scales of
$\sim$0.02\AA~. These reflect the imperfection in the model line
profiles, more Gaussian components reduce these residuals.

The $\zeta$Oph spectra were acquired at airmasses of 1.8 and 1.7 on
the 15$^\mathrm{th}$ and 16$^\mathrm{th}$ of June.  Both spectra show
absorption at 3958.15$\AA$ (at $3958.5~\AA$ in the observatory's rest
frame).  This is absent in the spectrum from 14$^\mathrm{th}$ of June
and in all other targets. Since all other spectra were acquired with
low airmass values compared with that of $\zeta$Oph on the
15$^\mathrm{th}$ and 16$^\mathrm{th}$, we conclude the 3958.15\AA~
feature is telluric. We did not attempt to include the 3958.15\AA~
feature in the fits. The telluric absorption propagates into the
residuals, and contributes to the noise level. Note that the telluric
feature has no effect on where we set the baseline level, because we
assigned zero weights to the spectral points in its neighbourhood.


Stahl et al. (\cite{sta89}) reported $R=77\pm 3$\footnote{note their
uncertainty is estimated as for $\sigma_1$ in this work, in the case
where all samples are assigned the same weight} from $\lambda$4232,
using a single Gaussian fit. But Stahl \& Wilson (\cite{sta92}) used
two Gaussian components to account for the broad wings detected by
Crane et al. (\cite{cra91}), and thus obtained a tighter fit to the
$\zeta$Oph CH$^+$ absorption, with $R=71\pm3$ for $\lambda$4232 and
$R=68 \pm 7$ for $\lambda$3957.

We also find that using two instead of one Gaussian components
significantly improves the fit, bringing reduced $\chi^2$ values for
the coadded $\zeta$Oph $\lambda$4232 spectrum from 2.07 for one
Gaussian (using the noise reference of the two Gaussian fit), to 1.05
for two Gaussians. But the resulting average for $R$ is unchanged, we
obtain $\langle R \rangle = 78.22 \pm 2.63 (3.75 )$ with one Gaussian,
$\langle R \rangle$ = 80.20$\pm 2.75 (4.19 )$ with two Gaussians.


%


What is the effect of including more than two Gaussians? It is
difficult to separate an error in the definition of the continuum
baseline from a real increase in confidence level of the fit. We
experimented on $\zeta$Oph with 3 and 4 Gaussians, keeping the noise
fixed to the rms dispersion of the residuals in the 2~Gaussian
fit. The reduced $\chi^2$ were respectively 0.92 and 0.88 for the 3
and 4 Gaussian fits to the spectral region centred on $\lambda$4232 in
the coadded spectrum of $\zeta$Oph. But some of the 4-Gaussians fits
to the $\zeta$~Oph spectra contained very narrow features,
corresponding to unphysical opacity components. Since we cannot
separate the uncertainty in the baseline definition, we adopt the
smallest number of Gaussians consistent with the dataset (with reduced
$\chi^2$ close to 1). We chose 3 Gaussians rather than 2 because
Crawford et al. \cite{cra94} report a 3-Gaussians decomposition of the
ultra-high resolution CH$^+$ profile toward $\zeta$Oph. The final
value from the UVES data is thus $\langle R \rangle = 78.47 \pm 2.65
(3.53 )$, with a 3-Gaussian fit.


All $R$ values for $\zeta$Oph are consistent within 2$\sigma$. The
value we obtain here is higher than previously reported by Stahl et
al. (\cite{sta92}) by 1.6~$\sigma$, adding the statistical errors in
quadrature.  For comparison Hawkins et al. (\cite{haw93}) give $R = 63
\pm 8$ for $\lambda$4232, $67\pm 19$ for $\lambda$3957, Crane et
al. (\cite{cra91}) give $R = 67.6 \pm 4.5$ for $\lambda$4232, and
Vanden Bout \& Snell (\cite{van80}) report $R =$~77$^{+17}_{-12}$,
also for $\lambda$4232.

The value reported in this article has an accuracy close to that of
Stahl et al. (\cite{sta92}), of $71\pm3$. Combining both measurement
with identical weights gives our best value for $\zeta$Oph as $R =
74.7 \pm 2.3$.

In what follows we will nonetheless use the UVES value of $R$ for
$\zeta$Oph to compare with the other lines of sight, $\langle R
\rangle  = 78.47 \pm 2.65 (3.53 )$ thus relying exclusively on
measurements obtained with the same instrument.


%

%

\subsubsection{HD110432}

The profile from Crawford (\cite{cra95}) is well fit by 1 Gaussian
only, but we obtain best results with 2 Gaussians.

The {\em Hipparcos} distance to HD110432 is 300$\pm$51~pc, which
confirms that the CH$^+$ absorption most likely arises in the
Coalsack, at a maximum distance of 180~pc (Franco \cite{fra89}).

The fit on the $\lambda$3957 line is excluded from the combined
measurement of $R$ because the adjacent telluric absorption feature
could affect the wings of the CH$^{+}$ absorption. Also the baseline
around $\lambda$3957 seems to be strongly affected by systematic
noise.  The $\lambda$3957 values differ from $\lambda$4232, which
otherwise give consistent numbers for each night.

\begin{figure*}
  \centering
  \includegraphics{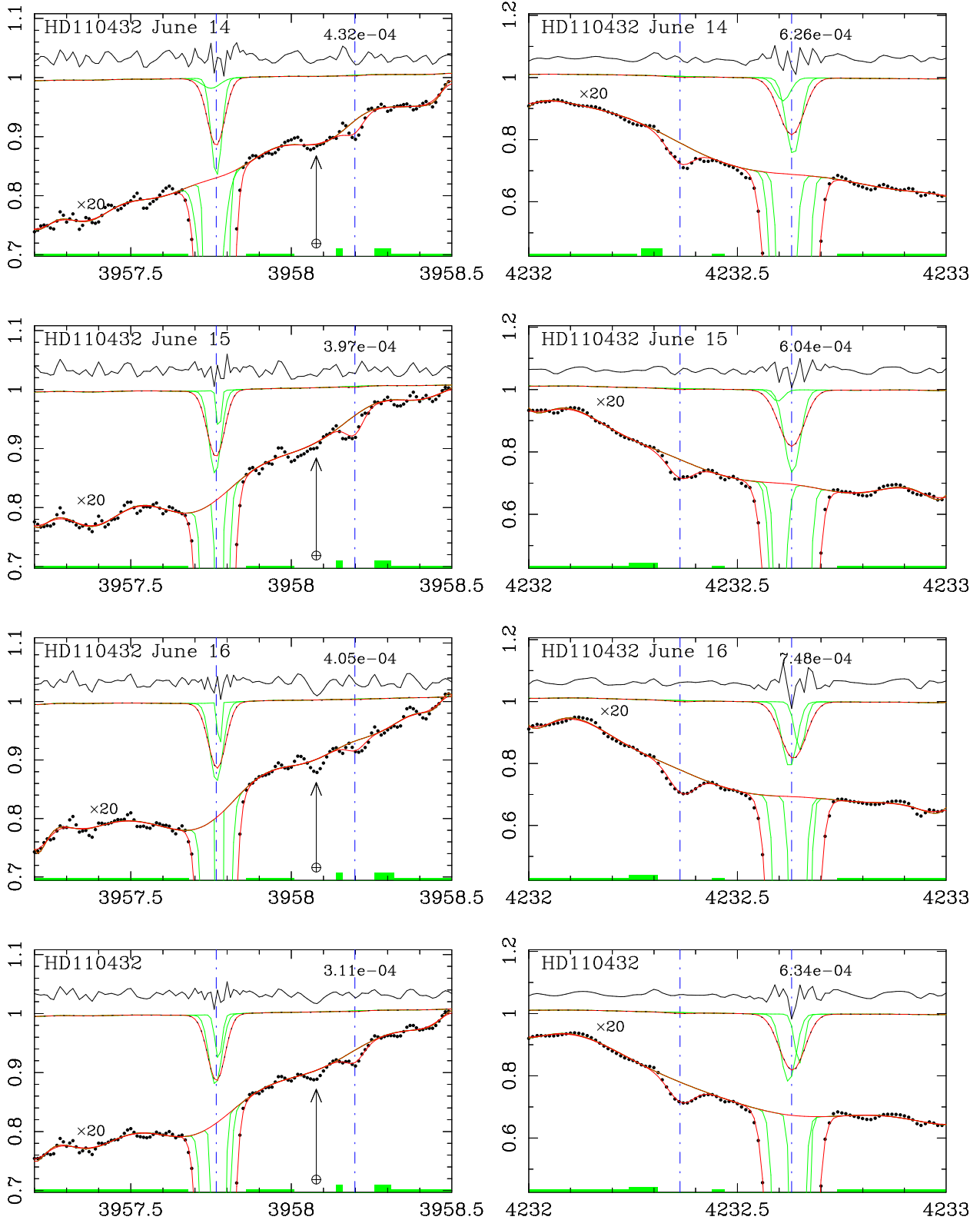}
  \caption{Same as Fig.~\ref{fig:zoph}. The fits on the $\lambda$3957
  lines are excluded from the combined measurement of $R$.  }
  \label{fig:hd110432}
\end{figure*}

%

\subsubsection{HD152235}

HD152235 is a member of the Sco~OB1 association, but as discussed by
Crawford (\cite{cra95}) the absorbing CH$^+$ lies in the Lupus
molecular cloud, at 170~pc (Murphy et al. \cite{mur86}).

\begin{figure*}
  \centering \includegraphics{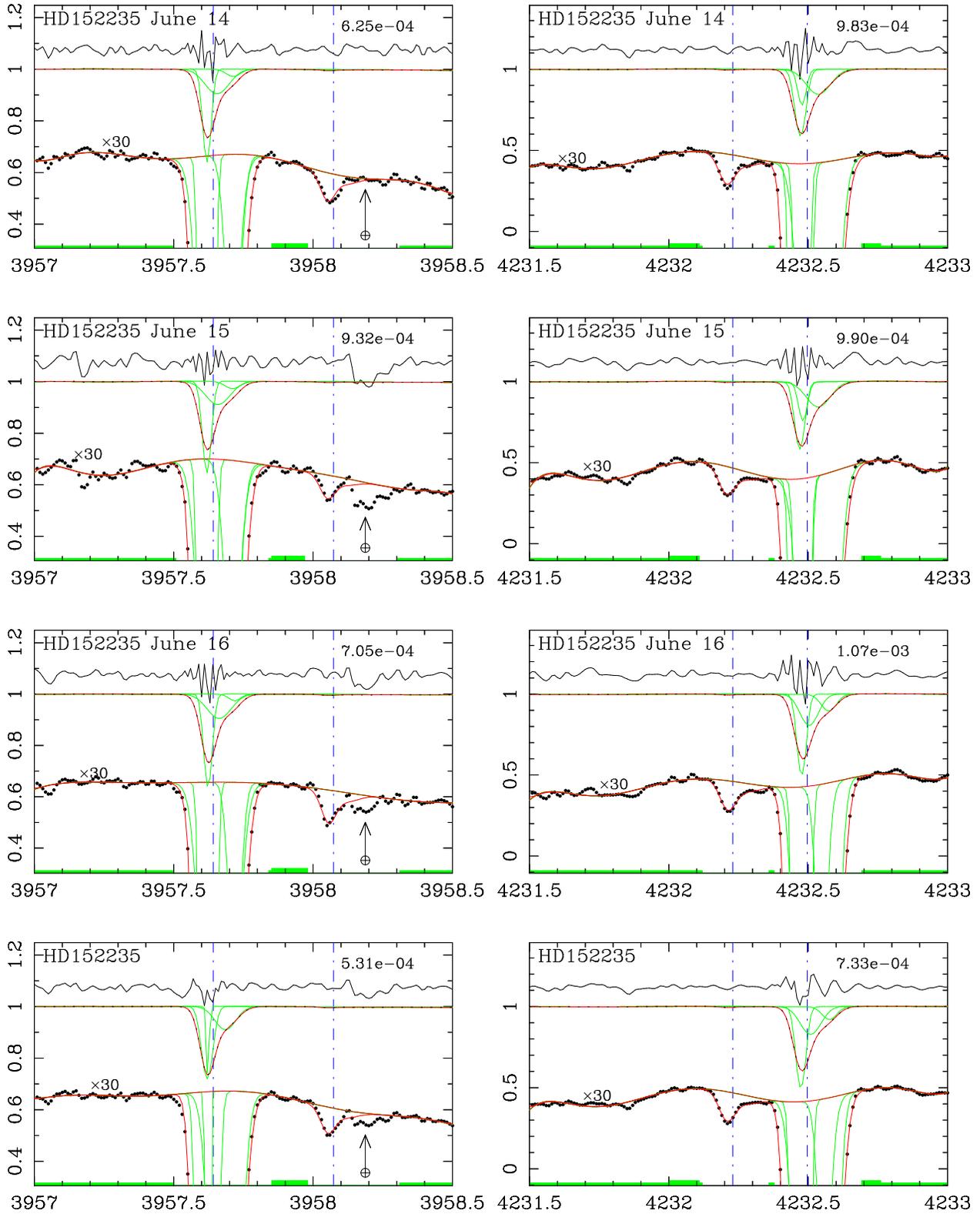}
  \caption{Same as Fig.~\ref{fig:zoph}. The fits on  $\lambda$3957 for
  June 15$^\mathrm{th}$ and 16$^\mathrm{th}$ are excluded from the
  combined measurement of $R$.  
  }
  \label{fig:HD152235}%
\end{figure*}

The profile from Crawford (\cite{cra95}) is fit by 3 Gaussians. Our
data, presented in Fig.~\ref{fig:HD152235},  also required 3 Gaussians
for $\lambda$3957.  


The CH$^{+}$ absorption has a red tail than difficults separating both
isotopes at 4232$\AA$, while the telluric absorption feature borders
the $^{13}$CH$^{+}$ line. It is difficult to pin down the underlying
continuum at 4232$\AA$. We tried two approaches to define the
continuum from edge to edge of the CH$^+$ absorption (from 4232$\AA$
to 4232.7$\AA$): a free fit (i.e. roughly a straight-line, with zero
weights over 4232$\AA$ -- 4232.7$\AA$), and the inclusion of a small
non-zero weight mid-way between the two isotopes (see
Fig.~\ref{fig:HD152235}).

The spectrum from 2003 June~14th is clear of telluric absorption (it
was observed with the smallest air mass, see Table~\ref{table:obs}),
so both overtones should give consistent values for $R$. This is why
we chose the baseline shown on Fig.~\ref{fig:HD152235}. The two
overtones give consistently high value for June~14th, and also on
average.  So we are rather confident of the ratio we report.  The fits
on $\lambda$3957 for June 15$^\mathrm{th}$ and 16$^\mathrm{th}$ are
excluded from the combined measurement of $R$ because they are
affected by telluric absorption under the red tail of $^{13}$CH$^+$.

Vladilo et al. (\cite{vla93}) reported $R = 126 \pm 29$. Our
measurement, which has better accuracy, gives lower value, although
still higher than the ISM average (see Sec.~\ref{sec:summary}).

This line of sight gives a significantly higher $R$ value than in
HD110432 at 4.2~$\sigma$, using the conservative uncertainties
$\sigma_2$, from the scatter of each individual measurement of $R$.


\subsubsection{HD152236}

\begin{figure*}
  \centering \includegraphics{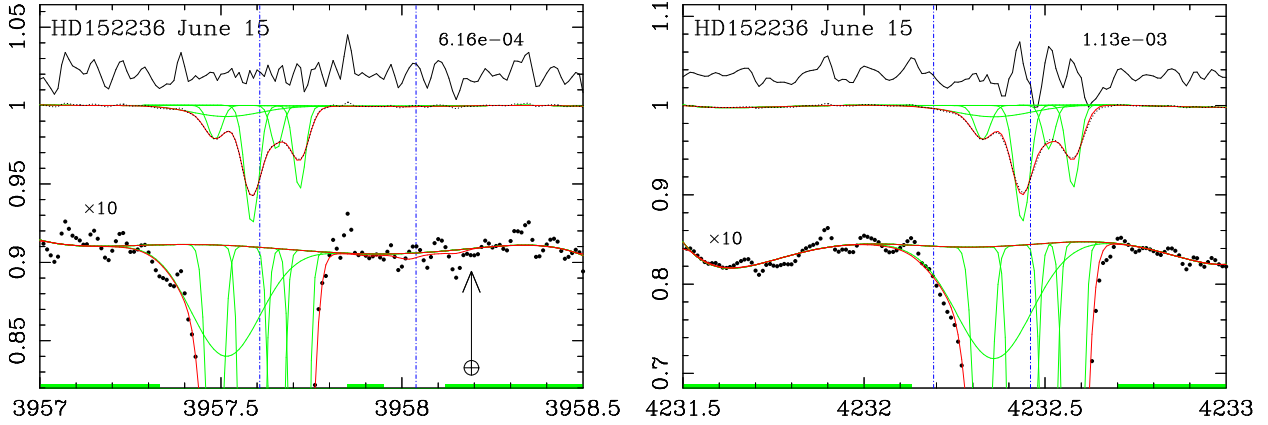}
  \caption{Same as Fig.~\ref{fig:HD154368}. No excess absorption is
  detected on the blue edge of $\lambda 4232 ~^{12}$CH$^+$. }
  \label{fig:HD152236}%
\end{figure*}

Fig.~\ref{fig:HD152236} presents the CH$^+$ spectra for HD152236,
along with our fits.  Crawford (\cite{cra95}) used 4 Gaussians, but we
needed 6.

HD152236, as HD152235, is a member of the Sco~OB1 association, so that
the absorption probably occurs in the Lupus cloud.

There is no manifest absorption at $\lambda$3957, but we give the
result of the formal fit, which allows assigning a lower limit $R$
value.  We take $f_\mathrm{uplimit}= \langle f \rangle + 3
\sigma_1(f)$, and $R > 1/ f_\mathrm{uplimit}$, giving $R > 61.4$ at
the 3~$\sigma$ level.

Although the $\lambda$4232 line has better signal to noise, both
isotopes are blended. We tried to apply the technique described below
in the case of HD154368 (Sec.~\ref{sec:hd154368}), but did not obtain
significant excess absorption in the blue wing of the main line. The
lower limit refers only to the $\lambda$3957 measurement. 



\subsubsection{HD152424}

HD152424, as HD152235, is a member of the Sco~OB1 association, so that
the absorption probably occurs in the Lupus cloud. 

Vladilo et al. \cite{vla93} report $R = 98 \pm 19$, but this target
gives us our lowest $R$ value, of $55.65 \pm 3.55 (15.61)$.

The two isotopes are blended at $\lambda$4232, so we apply the same
technique as for HD154368 (Sec.~\ref{sec:hd154368}). The residuals
under the $^{13}$CH$+$ absorption are consistent with the noise, which
gives us confidence in the inferred $R$ value. HD152424 provides our
best measurement of $R$ for blended lines, as obtained by keeping the
opacity profile fixed to the best fit on $\lambda$3957. But defining
consistent baselines between the two overtones is very difficult
without including the baseline as free parameter. So we decided to
ignore the significant residuals under the main isotope at
$\lambda$4232. We include the noisy residuals in the noise estimate,
so that the significance of the $R$ values derived from the blended
line is lowered. 

\begin{figure*}
  \centering
\includegraphics{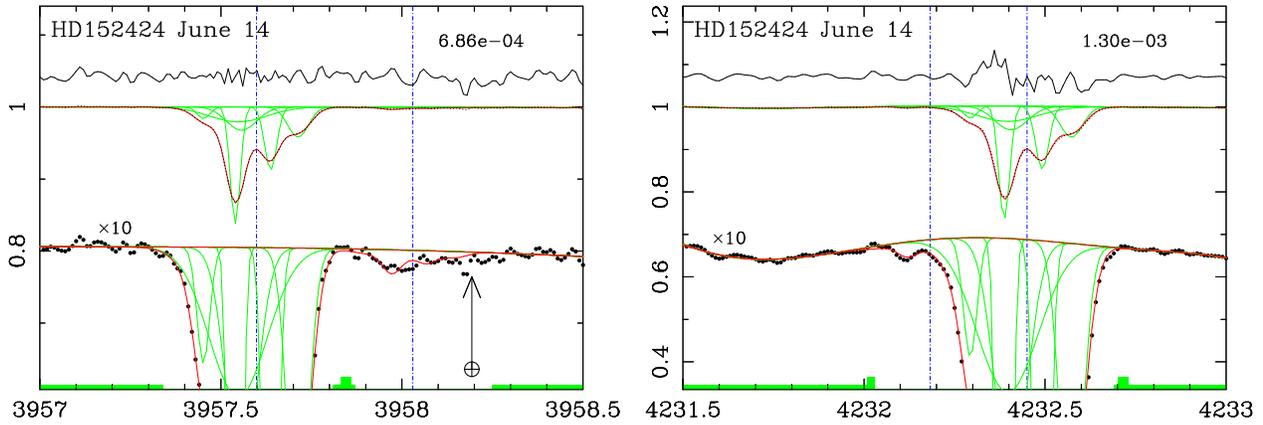}
  \caption{Same as Fig.~\ref{fig:HD154368}. $^{13}$CH$^+$ is visible
 as excess absorption on the blue edge of $\lambda 4232 ~^{12}$CH$^+$.}
  \label{fig:HD152424}%
\end{figure*}

The importance of quantifying systematic uncertainties is manifest in
HD152424: a visual inspection of Fig.~\ref{fig:HD152424} shows the fit
under $\lambda$3957 is very good, and that the baseline seems
smooth. However appeareances can be deceiving: the extrapolation of
the $\lambda$3957 fit to $\lambda$4232 gives a very different value of
$R$.  Had we considered only the $\sigma_1$ uncertainties, the value
for HD152424 would have been one of the most accurate. But because
CH$^+$ towards HD152424 is so broad the conservative uncertainty
$\sigma_2$ is much higher than $\sigma_1$ and reflects the systematic
uncertainty involved in the baseline definition.

\subsubsection{HD154368} \label{sec:hd154368}

This target was selected from the survey of Gredel et
al. (\cite{gre93}).

\begin{figure*}
  \centering
  \includegraphics{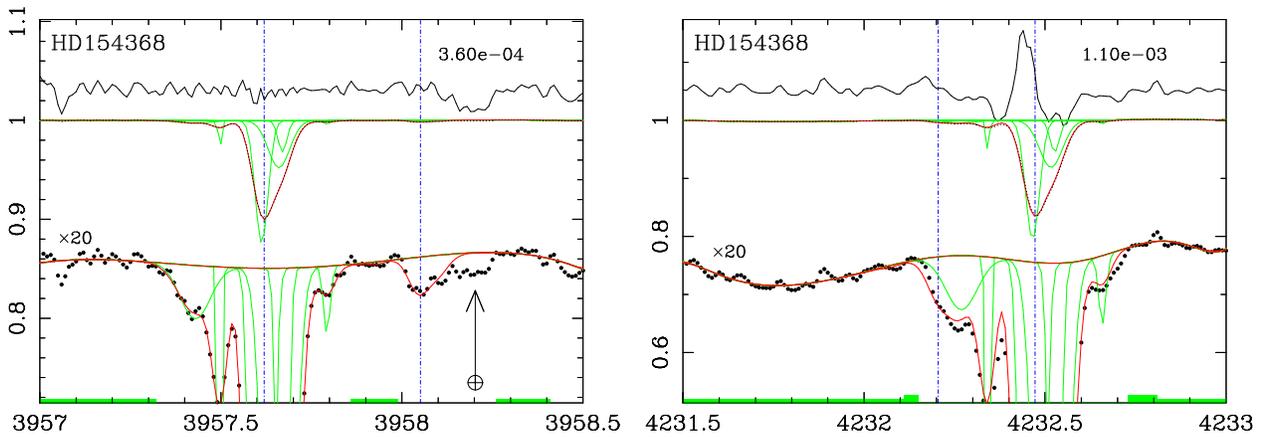}
  \caption{Same as Fig.~\ref{fig:zoph}. This line of sight has blended
  $^{12}$CH$^+$ and $^{13}$CH$^+$ absorption at $\lambda$4232. We
  extrapolate the $^{12}$CH$^+$ profile at $\lambda$3957 (see text for
  details). $^{13}$CH$^+$ is visible as excess absorption on the blue
  edge of $^{12}$CH$^+$. }
  \label{fig:HD154368}%
\end{figure*}

The CH$^+$ absoprtion for HD154368 (with an {\em Hipparcos} distance
of $364.8 \pm 128.3$~pc) shown on Fig.~\ref{fig:HD154368} has four
conspicuously distinct velocity components. The two isotopes are
manifestly blended for the $\lambda$4232 transition.

Our data cover the two overtones in the same UVES dichroic setting
simultaneously, and the spectral resolution is expected to be constant
with wavelength in such an instrumental setup. After checking we had
no detectable trend of varying resolution by measuring the arc line
widths, we attempted fitting $\lambda$4232 using the results of the
$\lambda$3957 fits.

A fit of both overtones simultaneously would be counterproductive
because the $\lambda$4232 region has better sensitivity, and would
dominate the fit, placing Gaussian components of $^{12}$CH$^+$ under
the rarer isotope.  Instead we kept the $^{12}$CH$^+$ opacity profile
as a function of velocity $\tau(v)$ as inferred from the $\lambda$3957
fit.  We then scaled $\tau(v)$ to $\tau(\lambda)$ for the
$\lambda$4232 region, which implies scaling component widths and
separations in wavelengths. After setting baselines as in the
unblended case, the optimization involved two free parameters, the
scaling factors on the $^{12}$CH$^+$ and $^{13}$CH$^+$ opacity
profiles.

The fits shown on Fig.~\ref{fig:HD154368} are rather poor in the case
of $\lambda$4232. Uncertainties in the baseline determination resulted
in significant residuals under the main isotope line. But these do not
affect the $^{13}$CH$^+$ absorption. We recompute the noise level from
the residuals, including the baseline uncertainties. The uncertainties
on $R$ tabulated in Table~\ref{table:all} are a result of this
exaggerated noise level.


\subsubsection{HD157038}

This line of sight also has blended absorption at $\lambda$4232, and
we apply the same technique as for HD154368. But the results on
$\lambda$4232 are useless because they are too sensitive on the
baseline definition, so we avoided the use of $\lambda$4232
altogether. 

Hawkins \& Meyer (\cite{haw89}, and references therein) give $R=38 \pm
12$ from $\lambda$4232 and use a distance of 1.7~kpc to HD157038. We
confirm this line of sight is enriched in $^{13}$C, with a lower $R$
value than the average ISM value (see Section~\ref{sec:ism}). 


\begin{figure*}
  \centering
\includegraphics{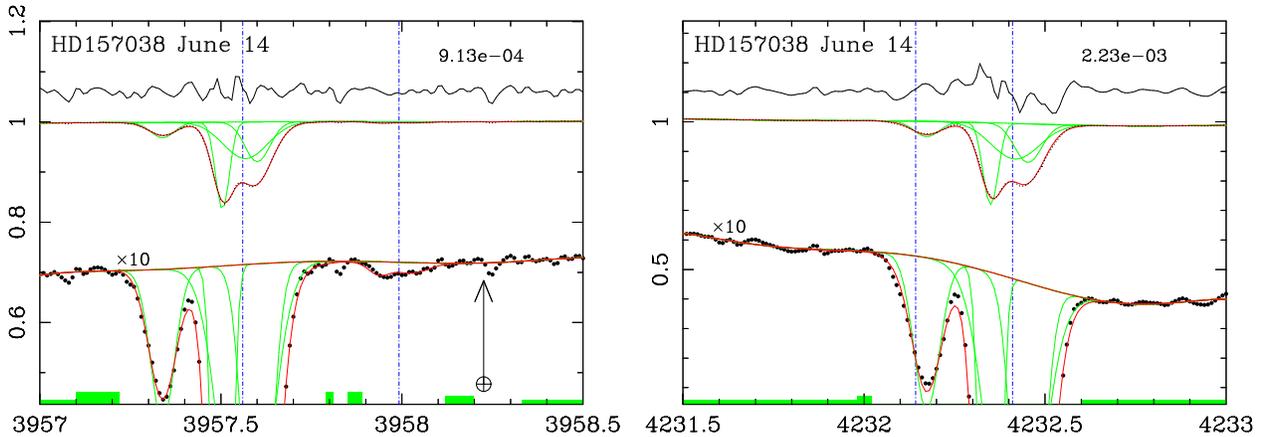}
  \caption{Same as Fig.~\ref{fig:HD154368}. The fit for $^{13}$CH$^+$
 absorption on the blue edge of $\lambda 4232 ~^{12}$CH$^+$ gives a 3~$\sigma$
 detection, but is barely visible on the Figure.}
  \label{fig:HD157038}%
\end{figure*}

\subsubsection{HD161056}

\begin{figure*}
  \centering
  \includegraphics{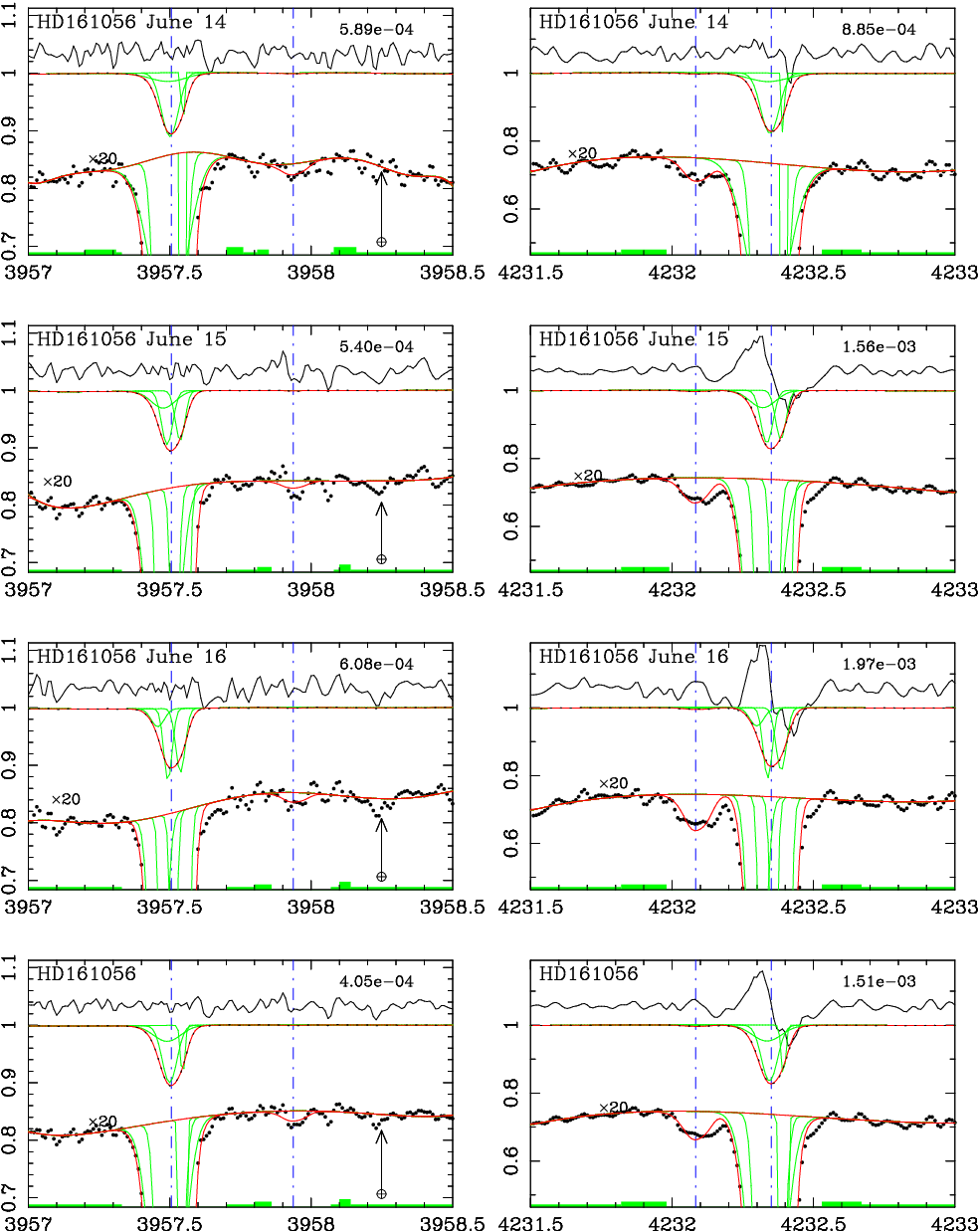}
  \caption{Same as Fig.~\ref{fig:zoph}. None of the $\lambda$4232
  spectra were used to combine individual measurements of $R$. There
  is excess absorption bridging the two isotopes at $\lambda$4232. The
  origin of the additional absorption, whether a telluric or a stellar
  feature, is unclear.}
  \label{fig:HD161056}%
\end{figure*}

$^{13}$CH$^+$ absorption towards HD161056 (selected from Gredel et
al. \cite{gre93}) is very weak at $\lambda$3957, and yet conspicuous
at $\lambda$4232 (see Fig.~\ref{fig:HD161056}). We first attempted to
fit the two overtones separately, and found widely discrepant values:
$f(3957) = 68.9\pm11.8(12.0)$, and $f(4232) = 159.3\pm 5.6(19.1)$. The
two isotopes may be blended at $\lambda$4232. We thus applied the same
technique as for HD154368, extrapolating the $\lambda$3957 opacity
profile to $\lambda$4232. But the inferred isotope ratios were still
very different (see Table~\ref{table:all}). There must be additional
absorption bridging the two overtones at $\lambda$4232, not due to
CH$^+$: the red tail of $\lambda$4232 is also seen at $\lambda$3957,
while the absorption at the blue edge of $\lambda$4232 is not seen at
$\lambda$3957. We cannot identify the nature of the additional
absorption, so we choose to exclude $\lambda$4232 from the combined
measurement of $R$ in HD161056.

The hipparcos distance to HD161056 is $425.7 \pm 145.5$~pc.

\subsubsection{HD169454}

This line of sight, selected from Gredel et al. (\cite{gre93}),
exibits at least 4 distinct velocity components. We use 6 Gaussians to
describe the profile, but did not attempt to compare reduced $\chi^2$
when varying the number of Gaussians because of the uncertainties in
the baseline definition.

The total number of free-parameters used to describe the main line is
6$\times$3, with 3 parameters per Gaussian.  The zero-absorption line
width for HD169454 is $\sim$0.8~{\AA}, or 80 spectral datapoints. But
the number of independent resolution elements under $^{12}$CH$^+$ is
about 16. Thus 18 free parameters may seem slightly excessive. But
over-constraining the opacity profile is of no consequence to the
derived isotopic ratio.

\begin{figure*}
  \centering
\includegraphics{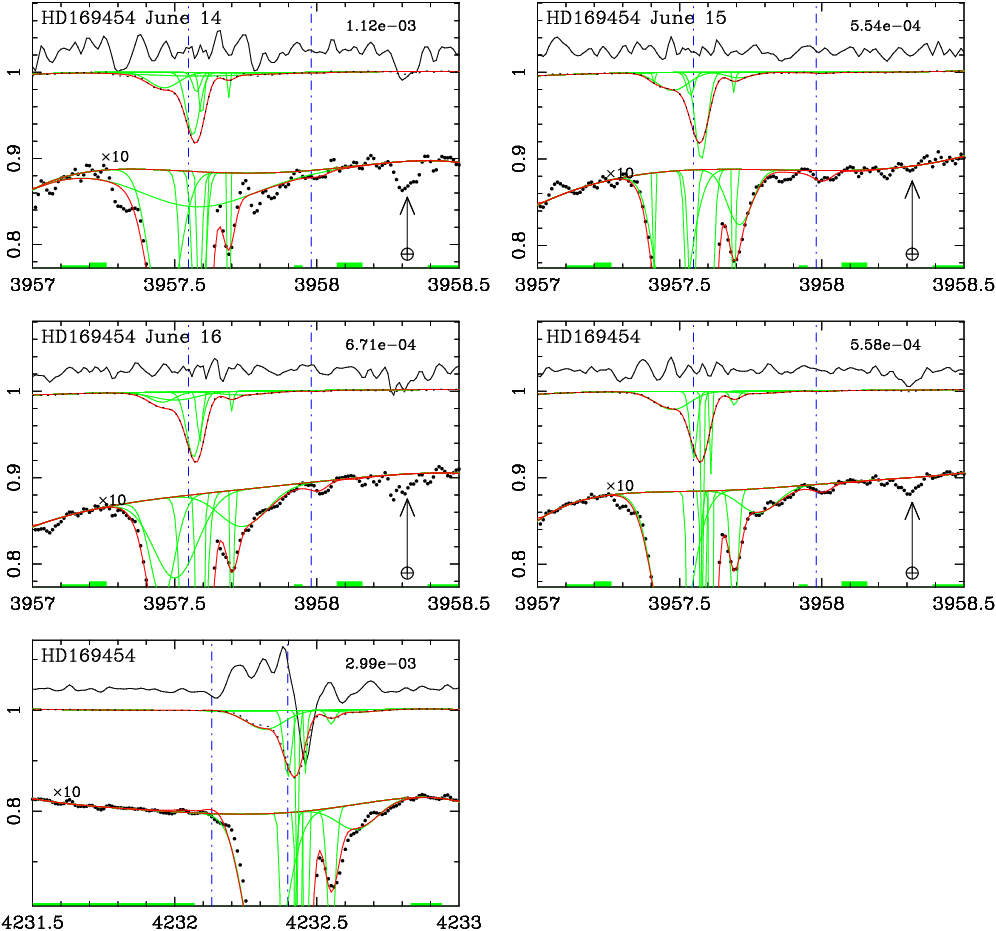}
  \caption{Same as Fig.~\ref{fig:HD154368}. The information from
  $\lambda$4232 is not used because $f$ is spuriously negative, and
  neither is that from $\lambda$3957 on June~14$^\mathrm{th}$ because
  of a non-ISM absorption feature at 3957.3$\AA$.  }
  \label{fig:HD169454}%
\end{figure*}

\begin{figure*}
  \centering \includegraphics{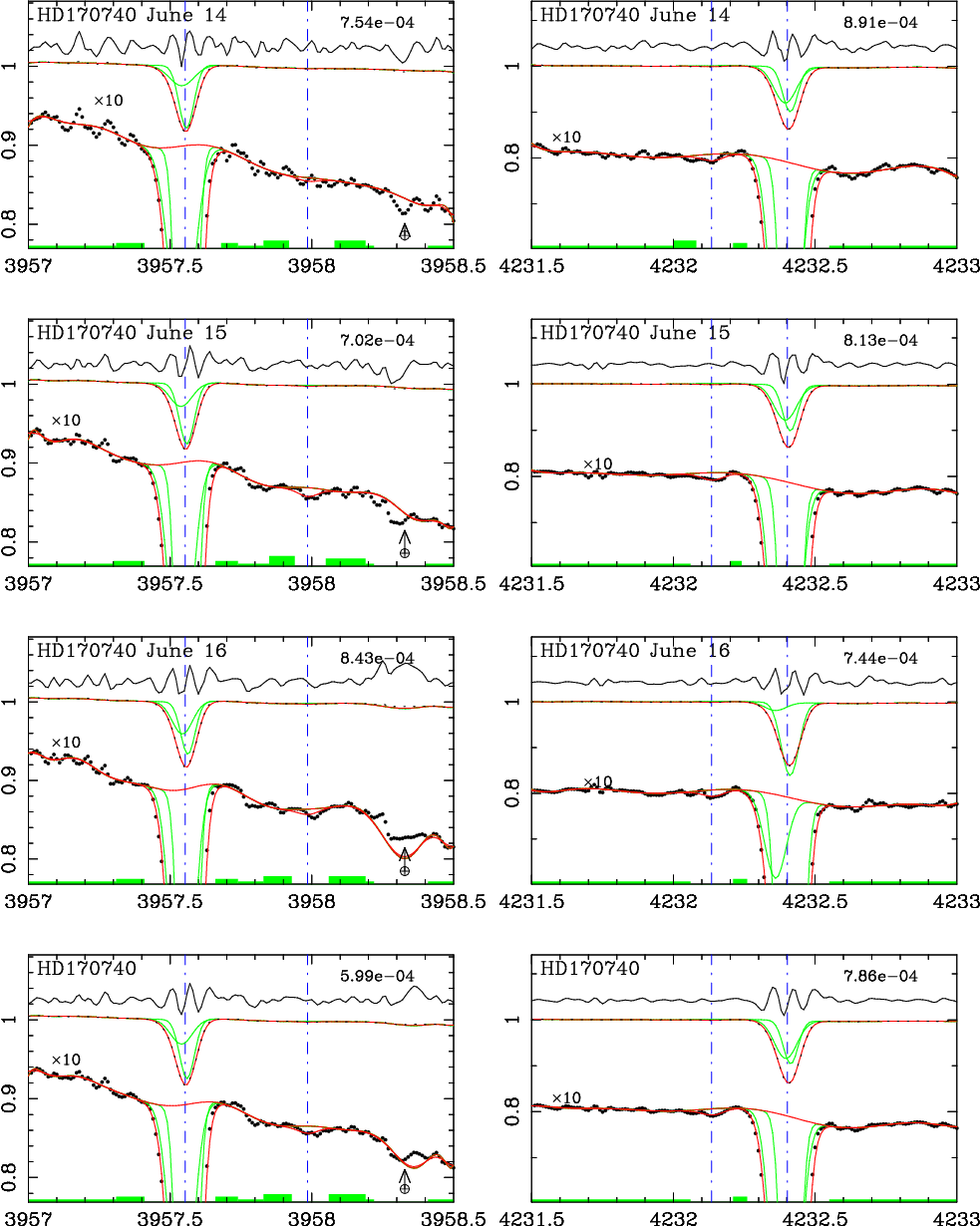}
  \caption{Same as Fig.~\ref{fig:HD154368}. The spectra around
    $\lambda$3957 have ragged baselines, presumably due to
    instrumental fringing. Only the coadded spectrum at $\lambda$3957
    is used for subsequent averaging of $R$ with that from
    $\lambda$4232.}
  \label{fig:HD170740}%
\end{figure*}

We use the same technique as for HD154368 to model the blended
$^{12}$CH$^+$ and $^{13}$CH$^+$ absorption at $\lambda$4232, but do
not obtain satisfactory results. The fits on $\lambda$4232 are shown
on Fig.~\ref{fig:HD169454} for completeness, and are not included in
the combined measurement of $R$ towards HD169454. An unphysical value
of $f$ is obtained, but this value is still consistent with zero
$^{13}$CH$^+$ within the uncertainties. Also excluded from the
combined measurement of $R$ is the fit to $\lambda$3957 from
June~14$^\mathrm{th}$ because it shows non-ISM absorption at
3957.3~$\AA$.  The $\lambda$3957.3 feature is not seen in other
spectra acquired at similar airmass (e.g. that of HD152235 on
June~15$^\mathrm{th}$), and is either an instrumental artifact or the
effect of passing clouds.

\subsubsection{HD170740}

This line of sight, Selected from Gredel et al. (\cite{gre93}), is
important since it has the highest $R$ value among those considered in
this work, together with HD161056.  The Hipparcos distance to HD170740
is $211.9 \pm 42.4 $~pc.

Unfortunately instrumental fringing, especially at $\lambda$3957,
limits the accuracy of the fits. The $\lambda$3957~ $^{13}$CH$^{+}$
line is only marginally detected in individual nights, but we give the
formal fits. The value we report relies on the nigthly spectra for
$\lambda$4232, and on the coadded spectrum for $\lambda$3957.


\section{Discussion} \label{sec:summary}

\subsection{Observed scatter and comparison with other sightlines} \label{sec:ism}

\begin{figure}
  \centering
  \resizebox{9cm}{!}{  \includegraphics{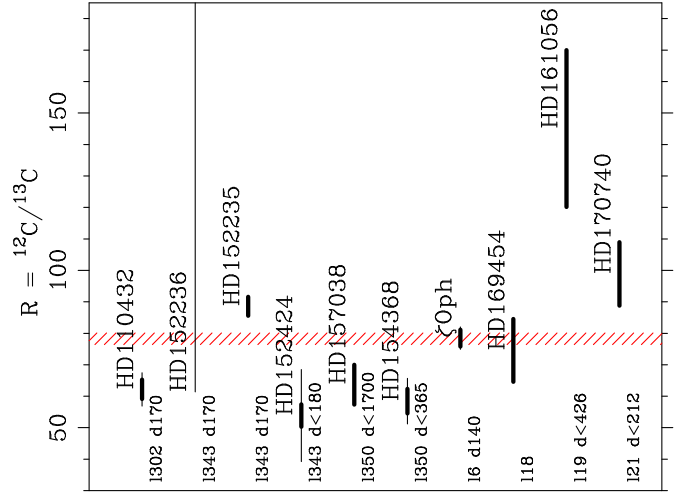} }
  \caption{A plot of individual $R$ values for the lines of sight
  studied in this work. The thick lines are $\pm \sigma_1$
  uncertainties, from the weighted average uncertainty of individual
  measurements. The thin lines are $\pm \sigma_2$ uncertainties, from
  the scatter of measurements for a given line of sight, and are set
  equal to $\sigma_1$ if $\sigma_2 < \sigma_1$. The hatched box is the
  weighted average of the measurements in the figure, and should be
  representative of the local ISM. This value is \rism.  The height of
  the hatched rectangle is $\pm 1~\sigma$. Galactic longitude and
  distance in parsec are indicated to the bottom right of each target. }
  \label{fig:all} 
\end{figure}

Figure~\ref{fig:all} summarises our results. The isotopic ratio $R$ is
lower for HD110432 than for HD152235 by 4.3~$\sigma$, and lower than
for HD170740 by 3.2~$\sigma$.  In this comparison we have used the
conservative uncertainties $\sigma_2$, derived from the scatter of $R$
for different nights and different overtones.

The weighted average of $R$ for the 9 lines of sight considered in
this work with $^{13}$CH$^+$ detections is $R_\mathrm{ISM} =
^{12}$C/$^{13}$C = \rism. The weighted 1~$\sigma$ dispersion of the
data is \scatterism.  The hypothesis that the observed scatter is
derived from a single value of $R$ can be tested by calculating the
corresponding value of $\chi^2/\nu$, where $\nu = 8$ is the number of
the degrees of freedom (the number of line of sights less one free
parameter). We obtain $\chi^2/\nu = 6.0$, which discards the
hypothesis at a confidence level of essentially 1.

The observed scatter is not due to the galactocentric abundance
gradient studied by Hawkins \& Meyer (\cite{haw89}), and predicted by
Galactic chemical evolution models. The targets in our sample all lie
within 30~deg of $l=0$, and at distances of at most $\sim$200~pc
(except for the clouds toward HD157038 and HD169454, whose distances
are unknown). Fig.~\ref{fig:all} also gives longitude and distance
information for each target.

Our value for $R_\mathrm{ISM}$ is significantly higher than those for
the Pleiades, $\xi$Per and P~Cyg.  Vanden Bout \& Snell (\cite{van80})
measured 49$^{+12}_{-8}$ and 59$^{+24}_{-13}$ for 20~Tau (Maia in the
Pleiades) and $\xi$Per.  Hawkins \& Jura (\cite{haw87}) report $R$
values of 46$\pm$15 for $\xi$Per, 40$\pm$9 for 20~Tau, 41$\pm$9 for
23~Tau (also in the Pleiades), and 45$\pm$10 for P~Cyg.  Hawkins et
al. (\cite{haw93}) confirm a low value of 49$\pm$15 for $\xi$Per.
Even taking 3~$\sigma$ errors, the measurements by Hawkins et al. are
below our results. One can ask whether the $^{13}$C enrichment in the
Pleiades is due to contamination from past stellar winds in the
cluster itself.

Aside from the $R$ values obtained by other groups and quoted so far
in this work, there are also those from Centurion \& Vladilo
(\cite{cen91}) for $\mu$Nor (67$\pm$6) and HD26676 (64$\pm$6), which
differ at 2~$\sigma$ with our average ISM value (including its
scatter).

\subsection{Are there D/H variations in the local ISM?}

From our data set, the values of $R$ vary by 7~$\sigma$ in the local
translucent clouds, where $\sigma$ is the uncertainty in the weighted
average value $\langle R_\mathrm{ISM} \rangle$.  This is strong
indication that the ISM at the same galactocentric distance is not
completely mixed. If this is the case in the chemical composition of
the ISM in general, then one expects that the D/H ratios in the local
ISM will also differ from place to place.  Moos et al. (\cite{moo02})
(see also Vidal-Madjar \cite{vid02}) summarise the {\em FUSE} results
towards nearby white dwarfs, located within 100~pc of the Sun.  The
D\,{\sc i}/H\,{\sc i} ratios measured with interstellar D\,{\sc i} and
H\,{\sc i} Lyman absorption varies by less than $\sim$10\% towards 7
white dwarfs within 100~pc of the Sun, at the limits of the Local
Bubble.  The fractional standard deviation of D\,{\sc i}/H\,{\sc i}
(the ratio of the scatter in the data to its weighted average) is
12\%, but cannot be distinguished from a single D\,{\sc i}/H\,{\sc i}
value within the uncertainties. By contrast, the D\,{\sc i}/H\,{\sc i}
ratios obtained with IMAPS show a factor of 2 scatter towards
$\gamma^2$\,Vel, $\zeta$\,Pup and $\delta$\,Ori, beyond the Local
Bubble, at distances of 300--500~pc (Jenkins et al. \cite{jen99},
Sonneborn et al. \cite{son00}).



Is the scatter in the observed interstellar D\,{\sc i}/H\,{\sc i} due
to variations in D/H? The primary objective of the IMAPS D\,{\sc
i}/H\,{\sc i} observations, as stated by Sonneborn et
al. (\cite{son00}), is ``to determine D/H with sufficient accuracy to
test for spatial inhomogeneity''. But the D/H ratio is notoriously
difficult to measure, and often involves the use of several different
instruments. The H\,{\sc i} lines are $>100$~km~s$^{-1}$ broad and
saturated so that N(H\,{\sc i}) is measured from the Ly\,$\alpha$
damping wings, while N(D\,{\sc i}) is best measured from Ly\,$\delta$
or Ly\,$\epsilon$ in the case of the IMAPS targets. The D/H ratio
involves carefully considering the effect of the underlying stellar
continua and the contribution from other interstellar H\,{\sc i}
components in the rising part of the curve of growth (with a high
ratio of equivalent width to opacity compared to the main absorption
line). Chemical fractionation and selective dissociation of HD affects
D\,{\sc i}/H\,{\sc i} for lines of sight where H$_2$ is observed, and
differential acceleration by radiation pressure could displace D\,{\sc
i} relative to H\,{\sc i} in the diffuse ISM. Bruston et
al. (\cite{bru81}) assign the D\,{\sc i}/H\,{\sc i} scatter known at
the time to such effects, and estimate a value of unperturbed D/H
almost a factor of two higher than the observed average D/H of
$1.5~10^{-5}$.

In addition Draine (\cite{dra04}) argues for variable depletion of D
onto dust grains, such that the gas phase D/H would be a function of
dust processing, which is in turn a function of Galactic
environment. The variations in D\,{\sc i}/H\,{\sc i} do not seem to be
a faithful measure of D/H.


\subsection{Inefficient mixing in the local ISM}

What is the level of elemental homogeneity in the ISM? What is the
power-spectrum of the dispersion in elemental abundance relative to
the total metalicity? Does heterogeneity increase with size? 

At least one extreme of spatial scale indicates perfect mixing: the
meteoritic evidence favours mixing at the molecular level of
interstellar dust from different origin, during the formation of the
solar system. Zinner et al.  (\cite{zin91}, their Table~1) measured
$^12$C/$^{13}$C values from 3 to 1000 in different SiC grains found in
the Murchison meteorite. At some point in the presolar cloud the SiC
grains must have been mixed from a previously heterogenous
distribution.

In a recent review of interstellar turbulence and mixing, Scalo \&
Elmegreen (\cite{sca04}) summarise current knowledge of the level of
elemental heterogeneity in the ISM. The stellar return to the ISM
should be spotty and poorly mixed. Scalo \& Elmegreen explain that in
contrast with diffusive processes, turbulence transport does not
homogenize the gas at the atomic level. Yet the heterogeneity is
notoriously difficult to detect.  The available gas-phase diagnostics
depend on local excitation, while the stellar data give upper limits
only on the dispersion of elemental composition.

But the $^{12}$CH$^+$/$^{13}$CH$^+$ ratio, in contrast to D\,{\sc
i}/H\,{\sc i}, is a particularly sensitive probe of the mixing
efficiency of the ISM. In terms of the fractional standard deviation
used by Moos et al. (\cite{moo02}), the scatter we have measured is
$\sigma_\mathrm{ISM}/\langle R_\mathrm{ISM} \rangle =$\fracdev on a
scale of 100~pc. Considering the precise location of the absorbing
material is unknown, we take the Moos et al. (\cite{moo02})
measurement of a 10\% scatter of D\,{\sc i}/H\,{\sc i} from the mean
value as a indication of inhomogeneity of the local ISM, over 100~pc
scales. This value is similar to the variations reported in this work,
even though D is burned in stellar interiors, rather than produced as
is the case of $^{13}$C. Although their absolute yields are different,
the mixing process (i.e. passive scalar turbulence, Scalo \& Elmegreen
\cite{sca04}) should be similar for both species.


%

\subsection{Is the solar value for $R$  comparable to ISM measurements?}

The terrestrial value of $R$ is 90 (Rosman \& Taylor \cite{ros98}),
and is usually extrapolated to the Sun. However, the preliminary
results of Ayres et al. (\cite{ayr05}) indicate a photospheric carbon
isotopic ratio of 70. The average value of our measurements of $R$ in
the local ISM, \rism, is slightly lower than the solar value of 90 (or
even higher than the photospheric value of 70). But in fact $\langle
R_\mathrm{ISM} \rangle$ is indistinguishable from the solar carbon
isotope ratio given the observed scatter of \scatterism.


Models of galactic chemical evolution predict an enhancement of
$^{13}$C relative to $^{12}$C with time. Thus our result that
$^{13}$C/$^{12}$C has essentially remained constant over the past
4.5~Gyr is surprising.  The inconsistency is worsened when comparing
with the preliminary results of Ayres et al. (\cite{ayr05}).  We may
be affected by a sampling bias. Around 30 lines of sight or more are
required to faithfully estimate $\langle R_\mathrm{ISM} \rangle$.

But is the solar value truly representative of the ISM at the time of
collapse of the solar nebula?  The process of CO fractionation
followed by condensation on cold dust grains may have increased
$^{13}$C in the pre-solar dust. With an enhanced dust-to-gas ratio
through sedimentation in the accretion disk of the Sun, the result may
have been a modification in the final $R$ value. 

We imagine a $^{13}$C-rich disk with a dust mass of 0.01~$M_\odot$,
dust to gas ratio of unity, accreting on a zero-age Sun with
metalicity Z=0.01. The $^{13}$C-poor atmosphere of the disk may have
been blown away by the early solar wind, so that once the accretion of
the disk is concluded, the Sun reaches $Z=0.02$ and higher
$^{13}$C/$^{12}$C than the pre-solar cloud.

Aside from the possibility of preferential accretion of $^{13}$C over
$^{12}$C, there is evidence that the solar nebula differed in
composition from the bulk ISM. For instance the metalicity of B stars
in Orion is lower than that of the Sun, even after 4.5~Gyr of Galactic
chemical evolution, indicating that the Sun must have been
exceptionally metal-rich. This discrepancy has been interpreted in the
framework of Galactic diffusion (Wielen \cite{wie77}), by which the
Sun diffused from its birthplace outwards in Galactocentric radius by
$\Delta r_\mathrm{GAL} = 1.9~$kpc.  Since $^{13}$C/$^{12}$C is also
predicted to increase with $r_\mathrm{GAL}$, Galactic diffusion may
bring in agreement the ISM and solar values of $R$ at the time of
solar birth (Wielen \& Wilson \cite{wie97}).

But perhaps the most compelling evidence that isotopic ratios in the
Sun cannot be taken as direct constraint on the bulk ISM are isotopic
anomalies in meteorites, such as the presence of extinct
radionuclides.  For instance Zinner et al. (\cite{zin91}) found that
$^{26}$Al/$^{27}$Al$\approx$1 in the pre-solar nebula. They measured
the concentration of the decay product of $^{26}$Al, $^{26}$Mg, locked
in SiC grains from the Murchison meteorite. But in other meteorites
$^{26}$Al/$^{27}$Al$\approx$1/4000 in Al-rich minerals from refractory
inclusions not representative of the pre-solar composition. Since the
half-life of $^{26}$Al is $10^6~$yr, the pre-solar enrichment in
$^{26}$Al suggests pollution by a nearby source of $^{26}$Al, such as
a supernova explosion (see Zinner et al. \cite{zin91} and Clayton
\cite{cla94} for thorough discussions).


\section{Conclusions} \label{sec:conc}

The VLT-UVES spectra of CH$^+$ absorption in the lines of sight toward
10 bright stars have allowed us to measure the carbon isotopic ratio
$R= ^{12}$C/$^{13}$C with unprecedented accuracy. We confirm previous
measurements of $R$ toward $\zeta$Oph, and obtain significant scatter
in the local ISM.

Our value for $\zeta$Oph is $R = 78.47 \pm 2.65 (3.53 )$, where the
second uncertainty in parenthesis is that obtained {\em a-posteriori},
from independent measurements, and independent baseline and absorption
fits.  Combining this value with that of Stahl et al. (\cite{sta92})
gives our best value for $\zeta$Oph, $R = 74.7 \pm 2.3 $.

Averaging our measurements for the 10 lines of sight gives a value
representative of the local ISM: $\langle R_\mathrm{ISM} \rangle
=$\rism, with a weighted rms dispersion of \scatterism.  The
dispersion in isotopic ratio is 7 times the uncertainty in $\langle
R_\mathrm{ISM} \rangle$ - we detect heterogeneity at 7~$\sigma$.

%

The observed scatter in $^{13}$C/$^{12}$C is the first significant
detection of heterogeneity in the isotopic composition of the local
ISM, and was obtained with a single instrument and a homogeneous
analysis, with a diagnostic independent of the local excitation and of
chemical fractionation. The observed variations in $R$ can be
extrapolated to an overall elemental heterogeneity, and reveal that
mixing in the ISM is not perfect.

The solar carbon isotope ratio of $90$ is undistinguishable from the
present-day ISM ratio, considering its intrinsic scatter, even after
4.5~Gyr of galactic evolution. This statement is not affected by
instrumental uncertainties, although a sampling bias could distort our
ISM value.  We question whether the solar carbon isotopic ratio is
equal to that of the presolar nebula.

A larger sample of stars, and new data towards the Pleiades, the
Taurus cloud and P~Cyg with 8~m aperture telescopes, will allow a
better estimate of the average ISM value, and an improved
understanding of the relationship between elemental heterogeneity and
spatial scale in the ISM.

\begin{acknowledgements}

We thank the referee, Roland Gredel, for critical readings and
constructive comments that improved the article.  S.C acknowledges
support from Fondecyt grant 1030805, and from the Chilean Center for
Astrophysics FONDAP 15010003.
\end{acknowledgements}

\end{document}

%% file: table_ALL_paper.tex
\begin{tabular}{lcrrrcrrr}
             & \multicolumn{4}{c}{$\lambda$3957} & \multicolumn{4}{c}{$\lambda$4232} \\
target, date  & f  & $\sigma(f)$ & $ n_g $ & $W_\lambda$ & f  & $\sigma(f)$ & $ n_g $ &  $W_\lambda$  \vspace{1ex}  \\ \hline 
$\zeta$Oph, June 14  & $ \star\,138^{+  8.9}_{- 12.1} $ & 10.4 & 3 &   13.72(    3) &  $ \star\,121^{+  9.8}_{-  8.7} $ & 9.18 & 3 & 23.40(    5)   \vspace{1ex} \\ 
$\zeta$Oph, June 15  & $ \star\,128^{+ 14.8}_{- 12.3} $ & 13.5 & 3 &   13.68(    4) &  $ \star\,126^{+  8.8}_{-  8.5} $ & 8.67 & 3 & 23.37(    5)   \vspace{1ex} \\ 
$\zeta$Oph, June 16  & $ \star\,132^{+ 13.6}_{- 13.8} $ & 13.7 & 3 &   13.73(    4) &  $ \star\,124^{+ 12.4}_{-  9.5} $ & 10.8 & 3 & 23.45(    6)   \vspace{1ex} \\ 
 $\zeta$Oph, coadd  & $ \,111^{+  4.4}_{-  9.7} $ & 6.51 & 3 &   13.74(    2) &   $ \,124^{+  4.5}_{-  5.2} $ & 4.82 & 3 & 23.42(    3)   \vspace{1ex} \\ 
Average & \multicolumn{4}{c}{ 133.67 $\pm 7.06  (7.06  )$} &      \multicolumn{4}{c}{ 123.73$\pm 5.44  (5.44  )$}  \\ 
       & \multicolumn{8}{c}{ $\langle f \rangle$ =  127.44$\pm 4.31  (5.73  )$  ~~ $\langle R \rangle$  =  78.47$\pm 2.65  (3.53  )$}   \vspace{1ex} \\ \hline 
  HD110432, June 14  & $ \,85.6^{+ 17.8}_{- 15.0} $ & 16.3 & 2 &    7.85(    3) &  $ \star\,162^{+ 17.3}_{- 10.3} $ & 13.2 & 2 & 13.82(    4)   \vspace{1ex} \\ 
HD110432, June 15  & $ \,134^{+ 18.0}_{- 12.7} $ & 15.1 & 2 &    7.65(    2) &  $ \star\,146^{+ 15.8}_{- 10.5} $ & 12.9 & 2 & 13.83(    5)   \vspace{1ex} \\ 
HD110432, June 16  & $ \,66.8^{+ 16.9}_{- 13.3} $ & 14.9 & 2 &    7.71(    2) &  $ \star\,181^{+ 22.6}_{- 10.7} $ & 15.6 & 2 & 13.86(    5)   \vspace{1ex} \\ 
 HD110432, coadd  & $ \,86.4^{+ 13.9}_{-  9.9} $ & 11.7 & 2 &    7.71(    2) &   $ \,159^{+ 16.4}_{- 11.3} $ & 13.6 & 2 & 13.64(    4)   \vspace{1ex} \\ 
Average       & \multicolumn{8}{c}{ $\langle f \rangle$ =  160.86$\pm 7.94  (13.76 )$  ~~ $\langle R \rangle$  =  62.17$\pm 3.07  (5.32  )$}   \vspace{1ex} \\ \hline 
  HD152235, June 14  & $ \star\,115^{+  7.0}_{-  9.9} $ & 8.32 & 3 &   24.81(    5) &  $ \star\,116^{+  8.0}_{-  6.7} $ & 7.32 & 3 & 41.14(    8)   \vspace{1ex} \\ 
HD152235, June 15  & $ \,96^{+ 11.0}_{- 14.1} $ & 12.4 & 3 &   25.03(    7) &  $ \star\,108^{+  7.8}_{-  7.0} $ & 7.38 & 3 & 40.99(    8)   \vspace{1ex} \\ 
HD152235, June 16  & $ \,135^{+  9.2}_{- 10.5} $ & 9.79 & 3 &   24.85(    5) &  $ \star\,113^{+  8.6}_{-  7.1} $ & 7.84 & 3 & 41.16(    8)   \vspace{1ex} \\ 
 HD152235, coadd  & $ \,106^{+  7.3}_{-  6.8} $ & 7.08 & 3 &   24.82(    4) &   $ \,114^{+  6.6}_{-  4.4} $ & 5.39 & 3 & 41.06(    6)   \vspace{1ex} \\ 
Average &   & &  &    \multicolumn{4}{c}{ 112.33$\pm 4.33  (4.33  )$}  \\ 
       & \multicolumn{8}{c}{ $\langle f \rangle$ =  112.90$\pm 3.84  (3.84  )$  ~~ $\langle R \rangle$  =  88.58$\pm 3.01  (3.01  )$}   \vspace{1ex} \\ \hline 
  HD152236, June 15  & $ \star\,62.1^{+ 32.1}_{- 35.2} $ & 33.6 & 5 &    9.67(    5) &  $ \,17.4^{+ 34.6}_{- 34.6} $ & 34.6 & 5 & 18.01(    9)   \vspace{1ex} \\ 
      &  \multicolumn{8}{c}{ ~~  ~~ $\langle R \rangle$ = $>$  61.39 }  \vspace{1ex} \\ \hline 
  HD152424, June 14  & $ \star\,233^{+ 17.0}_{- 16.4} $ & 16.7 & 6 &   20.39(    6) &  $ \star\,132^{+ 17.8}_{- 17.8} $ & 17.8 & 6 & 36.43(   11)   \vspace{1ex} \\ 
       & \multicolumn{8}{c}{ $\langle f \rangle$ =  185.72$\pm 12.18 (50.40 )$  ~~ $\langle R \rangle$  =  53.85$\pm 3.53  (14.61 )$}   \vspace{1ex} \\ \hline 
   HD154368, coadd  & $ \star\,184^{+ 14.0}_{- 12.8} $ & 13.4 & 6 &   10.64(    3) &   $ \star\,136^{+ 22.2}_{- 22.2} $ & 22.2 & 6 & 19.03(    9)   \vspace{1ex} \\ 
       & \multicolumn{8}{c}{ $\langle f \rangle$ =  171.18$\pm 11.47 (21.24 )$  ~~ $\langle R \rangle$  =  58.42$\pm 3.91  (7.25  )$}   \vspace{1ex} \\ \hline 
  HD157038, June 14  & $ \star\,157^{+ 15.0}_{- 16.2} $ & 15.6 & 4 &   28.21(    8) &  $ \,63.4^{+ 22.1}_{- 22.1} $ & 22.1 & 4 & 48.81(   19)   \vspace{1ex} \\ 
       & \multicolumn{8}{c}{    ~~~~~~~~ $\langle R \rangle$ =  63.69$\pm 6.33  $}   \vspace{1ex} \\ \hline 
  HD161056, June 14  & $ \star\,85.7^{+ 26.3}_{- 26.3} $ & 26.2 & 3 &   11.62(    4) &  $ \,182^{+ 16.3}_{- 16.3} $ & 16.3 & 3 & 20.51(    7)   \vspace{1ex} \\ 
HD161056, June 15  & $ \star\,56.4^{+ 16.8}_{- 18.7} $ & 17.6 & 3 &   11.28(    4) &  $ \,196^{+ 28.7}_{- 28.7} $ & 28.7 & 3 & 20.26(   11)   \vspace{1ex} \\ 
HD161056, June 16  & $ \star\,75.5^{+ 20.2}_{- 20.2} $ & 20.2 & 3 &   10.94(    4) &  $ \,281^{+ 36.2}_{- 36.2} $ & 36.2 & 3 & 20.07(   14)   \vspace{1ex} \\ 
 HD161056, coadd  & $ \,79.9^{+ 13.3}_{- 13.5} $ & 13.3 & 3 &   11.23(    3) &   $ \,220^{+ 27.7}_{- 27.7} $ & 27.7 & 3 & 20.26(   11)   \vspace{1ex} \\ 
Average       & \multicolumn{8}{c}{ $\langle f \rangle$ =   68.94$\pm 11.84 (11.97 )$  ~~ $\langle R \rangle$  = 145.05$\pm 24.91 (25.18 )$}   \vspace{1ex} \\ \hline 
  HD169454, June 14  & $ \,51.6^{+ 50.8}_{- 49.4} $ & 49.9 & 6 &   10.70(   11) &   &  &  &  \vspace{1ex} \\ 
HD169454, June 15  & $ \star\,141^{+ 20.6}_{- 25.4} $ & 22.8 & 6 &   10.23(    5) &   &  &  &  \vspace{1ex} \\ 
HD169454, June 16  & $ \star\,123^{+ 31.4}_{- 26.6} $ & 29 & 6 &   10.27(    6) &   &  &  &  \vspace{1ex} \\ 
 HD169454, coadd  & $ \,112^{+ 24.2}_{- 24.2} $ & 24.3 & 6 &   10.17(    5) &   $ \,-125^{+ 79.0}_{- 78.9} $ & 78.9 & 6 & 17.59(   28)   \vspace{1ex} \\ 
Average       & \multicolumn{8}{c}{ $\langle f \rangle$ =  134.12$\pm 17.92 (17.92 )$  ~~ $\langle R \rangle$  =  74.56$\pm 9.96  (9.96  )$}   \vspace{1ex} \\ \hline 
  HD170740, June 14  & $ \,50.8^{+ 32.9}_{- 33.8} $ & 33.3 & 2 &    7.98(    5) &  $ \star\,97.2^{+ 23.9}_{- 19.0} $ & 21.4 & 2 & 13.97(    7)   \vspace{1ex} \\ 
HD170740, June 15  & $ \,126^{+ 34.1}_{- 27.2} $ & 30.5 & 2 &    7.98(    5) &  $ \star\,110^{+ 21.1}_{- 16.2} $ & 18.5 & 2 & 13.91(    6)   \vspace{1ex} \\ 
HD170740, June 16  & $ \,77.2^{+ 38.7}_{- 34.8} $ & 36.7 & 2 &    7.82(    6) &  $ \star\,94^{+ 21.6}_{- 16.8} $ & 18.9 & 2 & 14.07(    5)   \vspace{1ex} \\ 
 HD170740, coadd  & $ \star\,103^{+ 28.0}_{- 23.8} $ & 25.9 & 2 &    7.85(    4) &   $ \,91.5^{+ 21.8}_{- 13.9} $ & 17.4 & 2 & 13.97(    6)   \vspace{1ex} \\ 
Average &   & &  &    \multicolumn{4}{c}{ 100.80$\pm 11.25 (11.25 )$}  \\ 
       & \multicolumn{8}{c}{ $\langle f \rangle$ =  101.15$\pm 10.32 (10.32 )$  ~~ $\langle R \rangle$  =  98.87$\pm 10.08 (10.08 )$}   \vspace{1ex} \\ \hline 
  \end{tabular} 